\documentclass[traditabstract]{aa}
\usepackage{psfrag,graphicx}
\usepackage{txfonts}
\usepackage{natbib}
\usepackage{ulem}
\usepackage{color}

\begin{document}
\bibliographystyle{aa}

\title{Star formation near an obscured AGN}
\subtitle{Variations in the initial mass function}
\author{S. Hocuk \inst{1}
       \and
       M. Spaans \inst{1}
       }
\institute{Kapteyn Astronomical Institute, University of Groningen,
          P.~O.~Box 800, 9700 AV Groningen \\
          \email{seyit@astro.rug.nl, spaans@astro.rug.nl}
         }
\titlerunning{Star formation in AGN}
\authorrunning{S.~Hocuk \& M.~Spaans}
\date{Received \today}

\abstract
{The conditions that affect the formation of stars in radiatively and mechanically active environments are quite different than the conditions that apply to our local interstellar neighborhood. In such galactic environments, a variety of feedback processes can play a significant role in shaping the initial mass function (IMF). Here, we present a numerical study on the effects of an accreting black hole and the influence of nearby massive stars to a collapsing, 800 $\rm M_{\odot}$, molecular cloud at 10 pc distance from the black hole. Our work focusses on the star-forming ISM in the centers of (U)LIRGS. We therefore assume that this region is enshrouded by gas and dust and that most of the UV and soft X-ray radiation from the BLR is attenuated along the line of sight to the model cloud. We then parametrize and study radiative feedback effects of hard X-rays emanating from the black hole broad line region, increased cosmic ray rates due to supernovae in starbursts, and strong UV radiation produced by nearby massive stars. We also investigate the importance of shear from the supermassive, $\rm 10^{6}-10^{8} ~M_{\odot}$, black hole as the star-forming cloud orbits around it. A grid of 42 models is created and calculated with the hydrodynamical code FLASH. We find that thermal pressure from X-rays compresses the cloud, which induces a high star-formation rate early on, but reduces the overall star-formation efficiency to about 7\% due to gas depletion by evaporation. We see that the turn-over mass of the IMF increases up to a factor of 2.3, $\rm M_{turn}=1-1.5 ~M_{\odot}$, for the model with the highest X-ray flux (160 $\rm erg ~s^{-1} ~cm^{-2}$), while the high-mass slope of the IMF becomes $\Gamma \gtrsim -1$ ($\rm \Gamma_{Salpeter}=-1.35$). This results in more high-mass stars and a non-Salpeter initial mass function. Cosmic rays penetrate deeply into the cloud and increase the gas temperature to about 50 K for rates that are roughly 100 times Galactic and 200 K for 3000 times Galactic, which leads to a reduced formation efficiency of low-mass stars. While the shape of the mass function is preserved, high cosmic ray rates increase the average mass of stars, thereby shifting the turn-over mass to higher values, i.e., up to several solar masses. Due to this process, the onset of star formation is also delayed. We find that UV radiation plays only a minor role. Since UV photons cannot penetrate a dense, n $\rm \geqslant 10^{5} ~cm^{-3}$, cloud deep enough, they only affect the late time accretion by heating the medium where the cloud is embedded in. When we increase the black hole mass, for a cloud that is at 10 pc distance, the turbulence caused by shearing effects reduces the star-formation efficiency slightly. Furthermore, shear weakens the effect of the other parameters on the slope of the IMF as well as the turn-over mass. The run with the most massive black hole, however, causes so much shear that the hydrodynamics is completely dominated by this effect and it severely inhibits star formation. We conclude that the initial mass function inside active galaxies is different than the one obtained from local environments. We also find that the combined effects of X-rays, cosmic rays, UV, and shear tend to drive toward a less pronounced deviation from a Salpeter IMF.}

\maketitle

\section{Introduction}
Star formation in extreme environments can be quite different than the formation of most stars in the Universe. In the inner kpc of galaxies, molecular clouds are exposed to intense radiation from active galactic nuclei (AGN) or starbursts \citep{1996ApJ...466..561M, 2005A&A...436..397M, 2009A&A...503..459P, 2010A&A...518L..42V}. Very close to the AGN, $\lesssim 0.1$ pc, gas collects into a massive AGN disc over some time, and as the disc becomes unstable, stars are able to form \citep{2006ApJ...643.1011P, 2007MNRAS.379...21N}. Slightly further away from the black hole, 1-100 pc, conditions are somewhat less extreme as the radiation is strongly attenuated by large columns of gas and dust. Such environments are typical of obscured AGN, as formed in (ultra-)luminous infrared galaxies ((U)LIRGS), with obscuring columns of 10$^{22}$-10$^{23.5}$ cm$^{-2}$ \citep{2005Ap&SS.295..143A, 2007A&A...476..177P, 2008A&A...488L...5L}. These regions have a strong impact on the initial phases of cloud evolution so that the final mass of stars or their formation efficiencies might drastically change. However, observing star formation in extreme environments is difficult. Results usually rely on indirect methods and are therefore often debated. The regions near AGN are also generally obscured \citep{2007ApJ...654L..49S, 2010A&A...523A...9H}. Predictions based on models and numerical simulations can aid observations to further our understanding of star formation. A good amount of numerical work has been done focussing on mechanical and radiative feedback effects in active galaxies and star-forming regions \citep{2001ApJ...556..837K, 2005ApJ...620..786K, 2008Sci...321.1060B, 2008ApJ...674..927A, 2008ApJ...675..188W, 2009MNRAS.394..191H, 2009ApJ...702...63W, 2010ApJ...713.1120K, 2010MNRAS.404L..79B, 2010A&A...510A.110H, 2010A&A...522A..24H, 2011ApJ...730...48P, 2011Sci...331.1040C, 2011MNRAS.413L..33L, 2011ApJ...737...63A, 2011MNRAS.412..469A}. Supported by numerical simulations, it is often thought that in these active regions, due to environmental conditions and feedback effects, the IMF should be different than the proposed universal mass function \citep{2007MNRAS.374L..29K, 2007MNRAS.379...21N, 2010ApJ...721.1531H, 2010ApJ...713.1120K, 2011MNRAS.413.2741G}.

The initial mass function is of fundamental importance for many areas of astrophysics. It is observed to behave like a power-law with a high-mass end that is well defined. The IMF is described as

\begin{equation}
\rm \frac{dN}{dM} \propto M^{-\alpha} \Longrightarrow \frac{dlogN}{dlogM} = -\alpha + 1 = \Gamma,
\label{eq:imf}
\end{equation}

\noindent
with N the number of stars in a range of mass dM, $\alpha$ the power-law index, and $\Gamma$ the slope above the characteristic mass of $\sim$0.3-0.5 M$_{\odot}$. First proposed by \cite{1955ApJ...121..161S}, a plethora of observations has led astronomers to believe that the shape might be universal in nature. Other astronomers have refined the shape of the distribution by especially improving on the low-mass end of the IMF \citep{1979ApJS...41..513M, 2001MNRAS.322..231K, 2003PASP..115..763C}. Until observations of extragalactic origin started to show variations in the IMF, most of the studies that claimed universality were coming from observations from our local neighborhood. Hints for deviation came from measurements of abundance patterns in extragalactic bulges \citep{2007A&A...467..117B, 2008A&A...478..335B}, enhancement of far infra-red luminosities in interacting galaxy systems \citep{2007MNRAS.377.1439B}, mass-to-light ratios of ultra-compact dwarf galaxies \citep{2009MNRAS.394.1529D}, and many others \citep{2005MNRAS.356.1191B, 2007ApJ...659..314P, 2008MNRAS.385..147D, 2008MNRAS.391..363W, 2008ApJ...674...29V, 2009eimw.confE..14E}. \cite{2010Natur.468..940V, 2011ApJ...735L..13V} found further evidence for variations in the IMF at the low-mass end from NaI and FeH band spectra in luminous elliptical galaxies. Still, strong evidence has yet to emerge.

Although the initial mass function may seem, theoretically, to be universal over a relatively wide range of environmental conditions \citep{2009arXiv0904.3302C, 2010ARA&A..48..339B, 2011arXiv1101.5172K}, perhaps even insensitive to small changes in metallicity \citep{2011ApJ...735...49M} and the Jeans mass \citep{2008ApJ...681..365E}, there are conditions that are far more extreme than the ones discussed in these papers, like in radiation dominated regions \citep{2005IAUS..235P..58M, 2006A&A...453..615P, 2010A&A...518L..42V, 2011A&A...525A.119M} and cosmic ray dominated regions \citep[CRDRs,][]{2011MNRAS.414.1705P}. Besides the importance of the thermodynamics for the Jeans mass, $\rm M_{\rm J}$, and thus the IMF, where $\rm M_{\rm J}$ is proportional to $\rm \rho^{-1/2} T^{3/2}$, the change in the equation of state is also essential. Assuming ideal gas conditions with a polytropic equation of state, $\rm P\propto\rho^{\gamma}$, where $\rm \gamma = 1 + d{\rm log}(T)/d{\rm log}(\rho)$, the softness of $\rm \gamma$ plays a major role, at a very early stage, in the fragmentation properties and the mass scale of unstable clouds \citep{2000ApJ...538..115S, 2003ApJ...592..975L, 2005IAUS..227..337K, 2005A&A...435..611J}.

The nuclei of active galaxies like Arp 220, Markarian 231, and even our Galactic center show signs of unusual star formation \citep{2005Natur.434..192F, 2005ASSL..327...89F, 2006ApJ...643.1011P, 2007MNRAS.374L..29K, 2009A&A...501..563E, 2009eimw.confE..14E, 2010ApJ...708..834B, 2009ApJ...693...56M, 2011A&A...525A.119M, 2011A&A...527A..36M}. In fact, even in the inner parsec of our Galaxy, i.e., Sgr A* and in M31, young stars are found at distances on the order of $\sim 0.03-0.3$ pc \citep{2003ApJ...594..812G, 2006ApJ...643.1011P, 2007MNRAS.374..515L}. All of the aforementioned places harbor a massive black hole. One can imagine that the conditions close to the black hole can indeed become quite extreme. Aside from strong gravity, accreting material onto a black hole will produce strong X-ray radiation (1-100 KeV). The dynamics of molecular clouds will be significantly affected by the irradiation of X-rays in X-ray dominated regions \citep[XDRs,][]{1996A&A...306L..21L, 1996ApJ...466..561M}. On the other hand, in starbursts, where star-formation rates can be a few hundred to a thousand solar masses per year \citep{1997ApJ...490L...5S, 1998Natur.394..241H}, UV radiation (6-13.6 eV) from O and B stars can be a significant presence and dominate the radiation field in these so-called photon dominated regions \citep[PDRs,][]{1999RvMP...71..173H}. However, where the gas is shielded from UV radiation, cosmic rays, created in supernova remnants or from winds in OB associations \citep{2008NewAR..52..427B}, will dictate the (minimum) temperature of the system, with energy densities of up to a few thousand times our galaxy \citep{2010ApJ...720..226P, 2011MNRAS.414.1705P, 2011A&A...525A.119M}. All of these environments have extremely different star-forming conditions, but will the stars that form out of them be much different?

In radiation dominated regions, the chemistry and thermal balance are determined by the radiation field \citep{2010A&A...513A...7S}. X-rays, in active galactic nuclei, are usually the dominant source for the excitation and chemistry of the inner disk out to a radius of $\rm \sim$160 pc, while UV radiation dominates the thermal balance in extreme starbursts and is generally important a bit further away from the accreting black hole \citep{2010A&A...518L..42V}. But there is compelling evidence that there is a strong link between AGN and starbursts \citep{2004ASPC..320..253S}. The question remains, how strongly star formation is affected by these extreme environments.

In an earlier numerical study, we showed that when the X-ray flux is as high as 160 $\rm erg ~s^{-1} ~cm^{-2}$, the stellar initial mass function of an 800 solar mass molecular cloud becomes top-heavy \citep{2010A&A...522A..24H}. This was a case where a molecular cloud orbiting at 10 pc from a $\rm 10^{7} ~M_{\odot}$ black hole under the impact of X-rays was compared to a cloud with the same conditions but in an X-ray free environment. This numerical study showed that under extreme conditions, the evolution of a molecular cloud and its stellar mass function will change, but it did not give insight into the quantitative details. Here, we present a parameter study on the influence of external radiation (X-rays, cosmic rays, and UV) and black hole shear on the IMF and the star-formation efficiency (SFE). In section \ref{sec:method} we introduce the numerical code FLASH and describe our additions to it. In section \ref{sec:conditions} we define the cloud models for all ambient conditions considered in the parameter study. We then present in section \ref{sec:results} our results on the effects of each condition for the evolution of the model clouds and show their phase diagrams, star-formation efficiencies, and initial mass functions. Finally, in section \ref{sec:conclusions}, we discuss the differences and similarities of these results in detail and present our conclusions.

\section{Computational method}
\label{sec:method}

\subsection{The numerical code}
The calculations in this work have been done using the hydrodynamical code FLASH 3 \citep{Dubey2009512}. For this study, we use the directionally split piecewise-parabolic method (PPM) which is described in detail in \cite{1984JCoPh..54..174C}. FLASH is well suited to handle these kinds of calculations as it is an adaptive mesh code and one that handles contact discontinuities very well. FLASH is provided with many and extensively tested modules that encompass a broad range of physics. Our simulation code is equipped with thermodynamics, hydrodynamics, (self-)gravity, multi-species, particles, and shocks, all from the standard modules of FLASH, as well as sink particles, radiative transfer, multi-scale turbulence, and refinement criteria (based on Jeans length and particles) that were added by us. The non-standard additions are explained in further detail in the following sections and in \cite{2010A&A...522A..24H}.

\subsection{Refinement criteria}
When one does a parameter study and has to perform many numerical simulations, saving time becomes crucial. In order to achieve this without suffering loss of quality, we made use of the adaptiveness of the FLASH code and wrote two independent refinement criteria that served our purpose. The simpler one of the two is based on sink particles. Since every sink particle accretes matter, and this matter can only be followed to within the sink particle's accretion radius, it is best to have the highest possible resolution here in order to resolve the affected volume properly. Particles can only be created in regions that have the highest grid resolution, however, they can move through the grid to unrefined regions. To this end, we simply say that wherever a sink particle is located, the grid must be refined to its maximum.

The second criterion is based on the Jeans length. In order to avoid numerical effects, like artificial fragmentation \citep{1997ApJ...489L.179T}, it is necessary to resolve the Jeans length, $\rm \lambda_{\rm J}$, in the simulation by at least 4 cells, $\rm \lambda_{\rm J} \geq 4 \triangle x$. Here, $\rm \triangle x$ is the size of the grid cell which is dependent on the resolution. In our experience, however, it is also likely that fragmentation can be artificially inhibited at even higher resolutions, up to twice the Truelove criterion, that is, 8 cells. In the presence of magnetic fields, even higher resolution constraints are found by \cite{2011ApJ...731...62F}. However, we do not consider magnetic field effects in this study. Therefore, we have chosen to resolve the Jeans length in our simulations by at least 10 grid cells, $\rm \ell_{\rm res,gas}=10$. This refinement criterion can be rewritten in the form of a density threshold and refines the grid if the following condition is met:

\begin{equation} \rm 
\rho \geq \frac {M_{\rm J}} {\frac{4}{3} \pi (\ell_{\rm res} \triangle x)^{3}},
\label{eq:jeansref}
\end{equation}
with the Jeans mass, $\rm M_{J}$, defined as:

\begin{equation} \rm 
M_{\rm J} = \frac{4}{3}\pi \lambda_{\rm J}^{3} \rho = \left( \frac {\pi c_{\rm s}^2} {G} \right)^{\frac{3}{2}} \rho^{-\frac{1}{2}}
\simeq \frac{90} {\mu^{2}} ~T^{\frac{3}{2}} ~n^{-\frac{1}{2}} ~(M_{\odot}),
\label{eq:jeans}
\end{equation}
as taken from \cite{2007A&A...475..263F}. Using this, Eq. (\ref{eq:jeansref}) can be further reduced to:

\begin{equation} \rm 
\rho \geq 1.455\times10^{15} \frac {T} {\mu (\ell_{\rm res} \triangle x)^{2}},
\label{eq:jeansref2}
\end{equation}
where $\rm \mu$ is the mean molecular weight, $\rm G$ is the gravitational constant, and T is the gas temperature.

Since we have a body in orbit, the Jeans refinement criterion will automatically follow the cloud in motion and increase the resolution of grid cells whenever required. The border resolution between the refinement levels is set-up in such a way that there is no sharp transition between the minimum and the maximum refinement. In order to be efficient, we also de-refine the regions where the Jeans length has stretched beyond 25 times the grid resolution. One can imagine this as a region where the cloud has just passed through.

\subsection{Multi-scale turbulence}
Turbulence is an important aspect of star formation. Never is the interstellar matter from which stars form fully homogeneous or kinematically quiescent. Typical velocity dispersions are found to be on the order of 1 km/s in most of the regions in our galaxy \citep{1981MNRAS.194..809L, 2001ApJ...555..178F, 2002ApJ...572..238C} and scale according to a power-law. It should be no surprise that gaseous clouds within active regions are more turbulent. Typical FWHM values are of the order of 5 km/s \citep{2009A&A...503..459P, 2011ApJ...731...41O}.

Numerical simulations have shown that turbulent strength, scaling, or type (compressible or solenoidal) can indeed be quite important and affect the results significantly \citep{2009ApJ...692..364F,2011MNRAS.413.2741G}. It is thus important to incorporate turbulence into the numerical code in a proper way. In all our simulations, we implement the turbulence using a Larson power-law, with a power spectrum $\rm P (k) \propto k^{-4}$ and thus $\rm \triangle v \propto \ell^{1/2}$, where $\rm k \propto \ell^{-1}$ is the scale length \citep{1981MNRAS.194..809L, 1999ApJ...522L.141M, 2004ApJ...615L..45H}. This is the predicted and observed behavior for compressible fluids. Since the grids in hydrodynamical simulations are discretized, and in block-structured grids the cell sizes usually increase with a factor of two for each resolution increment, the super-posed velocity decreases with the square-root of two for each higher level. In our case, the largest scale that we apply the multi-scale turbulence on is that of the cloud radius. Otherwise, the cloud as a whole would obtain a random motion. The smallest scale on which the turbulence is injected is determined by the maximum resolution at runtime, which is $\rm \triangle x = 1.76\times10^{16} \rm ~cm$ in our runs.

\subsection{The radiative transfer method}
Radiative transfer for PDRs and XDRs is handled through a radiation dominated region code written by \cite{2005A&A...436..397M}, with additional details in \cite{2007A&A...461..793M} and \cite{2008ApJ...678L...5S}. Pre-computed tables for gas temperature and chemical abundances are obtained from this code given an input of number density [$\rm cm^{-3}$], radiation flux [$\rm erg ~s^{-1} cm^{-2}$] for X-rays and UV, column density [$\rm cm^{-2}$], and metallicity. We choose solar abundances for all simulations. For the regions dominated by radiation, the code calculates all the heating processes (photo-ionization, yielding non-thermal electrons, FUV pumping followed by collisional de-excitation), cooling processes (atomic fine-structure and semi-forbidden lines), and molecular transitions (CO, H$_{2}$, H$_{2}$O, OH, and CH). Cosmic rays, dust-gas coupling, and secondary effects from X-rays  like internal UV, are treated as well. These tables are incorporated into FLASH. For the cloud models in the absence of radiation, we use isothermal conditions with an equation of state of the form $\rm P \propto \rho$ \citep{2010A&A...522A..24H, 2010A&A...510A.110H}.

We use a ray-tracing method to find the column densities during the simulation. At each time-step, the algorithm searches the grid and sums up the densities of each cell lying along the line of sight from the radiation source in order to find the total column. The cells are selected by determining whether the two angles of the cell edges with respect to the radiation source accommodate the angle of the target cell. Each cell is weighted with the length of the ray that passes through it. Once the column density is found, together with the density and the flux, the corresponding temperature is taken from the tables and used to update the variables in the simulation. See \cite{2011ApJ...730...48P} for additional details. In order to gain speed, the algorithm makes use of the block-structured mesh of FLASH by first searching the blocks that lie within the line of sight of the source, thereby only looking into the cells of those blocks. This makes the ray-tracing in effect about 500 times faster. This gain is welcome, since, by far, the most computational time is normally spent on finding the column densities. Additional speed is gained by not applying the ray-tracing algorithm to the whole grid at every timestep. We prioritize the regions that have densities above $\rm 100 ~cm^{-3}$ and update them regularly, but the lower density regions are only updated occasionally. These regions are of not much importance since stars cannot form here. Besides, the temperatures at low densities do not decrease quickly, since the cooling time scales as $\rm t_{\rm cool} \propto 1/n$ (optically thin and sub-thermal), and regular updates are not necessary.

The X-ray flux is an $\rm E^{-0.9}$ power-law between 1 and 100 keV. X-ray scattering is not very important, but is nonetheless treated in the XDR-code. Inverse Compton heating is not included, because our focus lies on the molecular gas ($<$1000 K) that is mostly heated by ionization of H and H$_2$, and by H$_3^+$ recombination \citep{2005A&A...436..397M}. A uniform background of cosmic rays prevents the temperature from dropping below 10 K. For this, a minimum cosmic ray ionization rate typical for the Milky Way, $\rm \zeta_{CR, Gal}=5\times10^{-17} \rm ~s^{-1}$, is assumed \citep{2000ApJ...538..115S}. The UV flux enjoys energies between 6 and 13.6 eV and follows the Habing spectrum \citep{1968BAN....19..421H}.

\subsection{Sink particles}
Sink particles are necessary if one wants to obtain a high dynamic range in density. These particles represent compact (proto-stellar) objects that are indivisible but can gain mass by accreting or merging. It is computationally very expensive to keep increasing the grid resolution in order to follow a molecular cloud collapse up to proto-stellar densities. We have created a sink particle algorithm for this purpose. There are several criteria that we check to establish irreversible collapse. When this stage is reached, one can stop following the collapse and can make a transition to sink particles. Following \cite{1995MNRAS.277..362B} and \cite{2010ApJ...713..269F}, with only small differences, we determine the point-of-no-return of a grid cell within a volume that is defined by an accretion radius $\rm r_{\rm acc}$ as: \\

\begin{enumerate}
 \item The grid cell is about to violate the Jeans criterion, Eq \ref{eq:jeansref2}.
 \item The grid cell and its neighbors are at the highest level of refinement.
 \item The grid cell has the deepest gravitational potential of all the cells within the volume.
 \item The divergence on each axis of the grid cell is negative, that is, $\rm d v_{i}/di < 0$, with $\rm i=\{x,y,z\}$, such that $\rm \triangledown \cdot \textbf{v} < 0$.
 \item The volume within the accretion radius is gravitationally bound, $\rm E_{\rm grav} + E_{\rm th} + E_{\rm kin} < 0$.
 \item The volume within the accretion radius is Jeans unstable, $\rm E_{\rm grav} + 2E_{\rm th} < 0$ and thus $\rm M > M_{\rm J}$.
 \item There are no other sink particles occupying the same cell.
\end{enumerate}

At the moment when a sink particle is created, material from within an accretion radius is taken away from the cells and put into the particle. We make sure that the total mass and momentum are conserved. The accretion radius is ideally smaller than the sonic radius, $\rm C_{\rm s}dt$, but one also wants to resolve this region well. On the other hand, one wants to keep the radius small, since one changes the physics within this radius. In practice we found that 2 cells was a reasonable compromise between these competing demands, also see \cite{2004ApJ...611..399K, 2010ApJ...713..269F}. Our accretion radius is $\rm r_{\rm acc}=3.5\times10^{16} \rm ~cm$ and reproduces the observed Chabrier IMF for runs M01 and M04, see section \ref{sec:results}.

The mass a sink particle obtains at creation is determined by a density threshold that is set by the maximum resolution. In this case, we follow the same resolution criterion as previously mentioned \citep{1997ApJ...489L.179T}, but let the code run as long as possible on gas dynamics before making the transition to sink particles. Using Eqs. \ref{eq:jeansref} and \ref{eq:jeansref2}, and taking $\rm \ell_{\rm res,sink}=6$, we determine how much material within $\rm r_{\rm acc}$ from the target cell is in excess of the threshold density and add this to the mass of the particle. Sink particles continue to accrete matter after they are created. Accretion onto particles is handled by a Bondi-Hoyle type of accretion \citep{1944MNRAS.104..273B, 1952MNRAS.112..195B, 1994ApJ...427..351R, 2004ApJ...611..399K}. This kind of accretion applies to a homogeneous flow of matter. Accretion increases with protostellar mass $\rm M$, but drops with decreasing ambient density $\rm \rho_{\infty}$ and higher Mach numbers $\rm \mathcal{M}$. The accretion rate is given by

\begin{equation} \rm
\dot{M} = 4\pi\rho_{\infty}G^{2}M^{2} \sqrt{ \frac {\lambda^{2}c_{\infty}^{2}+v_{\infty}^{2}} {\left(c_{\infty}^{2}+v_{\infty}^{2}\right)^{4}} }
=
\frac {4\pi\rho_{\infty}G^{2}M^{2}} {c_{\infty}^{3}} \sqrt{ \frac {\lambda^{2}+\mathcal{M}^{2}} {\left(1+\mathcal{M}^{2}\right)^{4}} },
\label{eq:bondi-hoyle}
\end{equation}
where $\rm c_{\infty}$ and $\rm v_{\infty}$ are the sound speed and the velocity of the gas far from the sink particle. Since there is no obvious choice for $\rm c_{\infty}$ and $\rm v_{\infty}$ in an inhomogeneous environment, we take the average value of the cells inside $\rm r_{\rm acc}$ as an alternative. $\rm \lambda$ is a non-dimensional parameter that depends on the equation of state of the gas and is on the order of unity. Throughout this work, we have adopted the value for an isothermal gas $\rm \lambda = \rm exp(3/2)/4 \simeq 1.12$ \citep{1952MNRAS.112..195B}.

Using the effective radius as first described by \cite{1952MNRAS.112..195B}

\begin{equation} \rm
r_{\rm BH} = \frac {GM} {c_{\infty}^{2}+v_{\infty}^{2}},
\label{eq:bondiradius}
\end{equation}
Eq. \ref{eq:bondi-hoyle} can be further reduced to

\begin{equation} \rm
\dot{M} = 4\pi\rho_{\infty}r_{\rm BH}^{2}c_{\infty} \sqrt{ \lambda^{2}+\mathcal{M}^{2} },
\label{eq:bondi-hoyle2}
\end{equation}
where $\rm \rho_{\infty}$ is defined as

\begin{equation} \rm
\rho_{\infty} = \frac {\bar{\rho}} {\alpha}.
\label{eq:rhoinf}
\end{equation}
Here, $\rm \bar{\rho}$ is the mean density within the accretion radius and $\rm \alpha$ is the density profile that depends on the cell size and the Bondi-radius. For the density profile, we use an exponential of the form $\rm \alpha = exp(r_{BH}/1.2\triangle x)$. We found this expression to behave well in the regime $\rm \triangle x \gtrsim r_{BH}$, where we usually are in, and equals to unity when $\rm \triangle x \gg r_{\rm BH}$. The factor of 1.2 is adopted from \cite{2004ApJ...611..399K} and should give good results in the range where $\rm \triangle x \sim r_{\rm BH}$.

Stars are often found in binaries, however, we cannot resolve binaries in our numerical code as our resolution is on the order of 2000 AU. Instead, we allow them to merge. These mergers do not affect our conclusions, see section \ref{sec:conclusions}. Binary formation or mergers should occur more frequently as the system virializes. Sink particles are eligible to merge when they are within each other's gravitational pull. We let an algorithm check for three conditions between every two particles, and when they pass these, we allow them to merge. This happens when:

\begin{enumerate}
\item The velocity difference between two particles, i and j, is less than the escape velocity between them, $\rm \triangle v_{ij} < \sqrt{\rm 2GM_{i}/\triangle r_{ij}}$, with $\rm M_{i}>M_{j}$.
\item The merging time is shorter than the hydrodynamical timestep, $\rm \pi \triangle r_{ij}^{3/2} / \sqrt{\rm 8G(M_{i}+M_{j})} < dt_{\rm hydro}$.
\item The forces of the other bodies, $\rm F_{\rm tot}$, are no longer significant, i.e., $\rm F_{\rm tot} < 0.05 \times F_{ij}$, where $\rm F_{ij} = GM_{i}M_{j}/\triangle r_{ij}^{2}$.
\end{enumerate}

\section{Models and initial conditions}
\label{sec:conditions}

\subsection{A grid of models}
For this work, a grid that consists of 42 models is computed which covers a range in the 4 parameters that we investigate. As stated before, the parameters that we vary are the X-ray flux, the cosmic ray rate, black hole mass, and the UV flux. A 10$^{7}$ M$_{\odot}$ black hole accreting at 10\% Eddington and at a distance of 10 pc, would radiate with a total flux of $\rm F_{Edd} = 10^4 ~(M_{bh}/10^{7}M_{\odot}) ~erg ~s^{-1} ~cm^{-2}$. We assume, since we are interested in (U)LIRGS, that there is a large absorbing column with $\rm \tau_{1keV} \sim 5$ and $\rm \tau_{UV} \gtrsim 50$ between our single star-forming cloud and the black hole BLR. Varying the black hole mass $\rm M_{\rm bh}$ or its distance from the model cloud $\rm d_{\rm bh}$, gives the same insight into the effects of gravitational shear, where the largest velocity difference $\rm \triangle v$ across the cloud depends on $\rm M_{\rm bh}/d_{\rm bh}^{3}$ for a given model cloud size much smaller than $\rm d_{bh}$. We have chosen to fix the cloud distance to the black hole and vary the shear using the black hole mass.

We study four X-ray fluxes by varying the Eddington rates, these are: 0, 5.1, 28, 160 $\rm erg ~s^{-1} ~cm^{-2}$, with roughly a factor of 5 between them. X-ray fluxes much higher than 100 erg s$^{-1}$ cm$^{-2}$ would completely ionize any molecular cloud of $\lesssim$ 10$^{6}$ cm$^{-3}$ and inhibit star formation. We take three black hole masses: $\rm 10^6$, $\rm 10^7$, and $\rm 10^8$ $\rm M_{\odot}$, which, for a fixed distance of $\rm d_{\rm bh}=10 \rm ~pc$, represents strong shear, medium shear, and negligible shear, respectively. Furthermore, we investigate three different cosmic ray rates: 1, 100, and 3000 times Galactic \citep[$\rm \zeta_{\rm CR, Gal}=5\times10^{-17} \rm ~s^{-1}$,][]{2000ApJ...538..115S}. Finally, we use two UV fluxes in our simulations for which we take: 0 and $\rm 10^{2.5} ~G_{0}$, with $\rm G_{0}=1.6 \times 10^{-3} \rm ~erg ~s^{-1} ~cm^{-2}$ \citep{1968BAN....19..421H}. For each parameter that we change, we keep all other conditions fixed. Almost all possible combinations between these parameters, with the exception of UV, is modeled in this study. Our fiducial model M01 has similar conditions to those of the Milky Way, albeit, with added gravitational shear. In this respect, model M04, which has negligible shear, is more closely related to the conditions of the solar neighbourhood. Table \ref{tab:table1} shows the parameter details of all models.

\begin{table*}[htb]
\begin{center}
\caption{Parameter details for each model}
\begin{tabular}{cccccccccc}
\hline
\hline
Model	& $\rm F_{X}$			& $\rm M_{\rm bh}$	& $\rm \zeta_{\rm CR}$	& $\rm F_{UV}$	&
Model	& $\rm F_{X}$			& $\rm M_{\rm bh}$	& $\rm \zeta_{\rm CR}$	& $\rm F_{UV}$	\\
	& $\rm [erg ~s^{-1} ~cm^{-2}]$		& $\rm [M_{\odot}]$	& [$\times$Galactic]	& $\rm [G_{0}]$	&
	& $\rm [erg ~s^{-1} ~cm^{-2}]$		& $\rm [M_{\odot}]$	& [$\times$Galactic]	& $\rm [G_{0}]$	\\
\hline
M01	& 0					& 10$^{7}$		& 1			& 0		&
M22	& 160					& 10$^{6}$		& 1			& 0		\\
M02	& 0					& 10$^{7}$		& 100			& 0		&
M23	& 160					& 10$^{6}$		& 100			& 0		\\
M03	& 0					& 10$^{7}$		& 3000			& 0		&
M24	& 160					& 10$^{6}$		& 3000			& 0		\\
M04	& 0					& 10$^{6}$		& 1			& 0		&
M25	& 0					& 10$^{8}$		& 1			& 0		\\
M05	& 0					& 10$^{6}$		& 100			& 0		&
M26	& 0					& 10$^{8}$		& 100			& 0		\\
M06	& 0					& 10$^{6}$		& 3000			& 0		&
M27	& 0					& 10$^{8}$		& 3000			& 0		\\
M07	& 5.1					& 10$^{7}$		& 1			& 0		&
M28	& 160					& 10$^{8}$		& 1			& 0		\\
M08	& 5.1					& 10$^{7}$		& 100			& 0		&
M29	& 160					& 10$^{8}$		& 100			& 0		\\
M09	& 5.1					& 10$^{7}$		& 3000			& 0		&
M30	& 160					& 10$^{8}$		& 3000			& 0		\\
M10	& 5.1					& 10$^{6}$		& 1			& 0		&
M31	& 0					& 10$^{7}$		& 1			& 10$^{2.5}$	\\
M11	& 5.1					& 10$^{6}$		& 100			& 0		&
M32	& 0					& 10$^{7}$		& 100			& 10$^{2.5}$	\\
M12	& 5.1					& 10$^{6}$		& 3000			& 0		&
M33	& 0					& 10$^{7}$		& 3000			& 10$^{2.5}$	\\
M13	& 28					& 10$^{7}$		& 1			& 0		&
M34	& 5.1					& 10$^{7}$		& 1			& 10$^{2.5}$	\\
M14	& 28					& 10$^{7}$		& 100			& 0		&
M35	& 5.1					& 10$^{7}$		& 100			& 10$^{2.5}$	\\
M15	& 28					& 10$^{7}$		& 3000			& 0		&
M36	& 5.1					& 10$^{7}$		& 3000			& 10$^{2.5}$	\\
M16	& 28					& 10$^{6}$		& 1			& 0		&
M37	& 28					& 10$^{7}$		& 1			& 10$^{2.5}$	\\
M17	& 28					& 10$^{6}$		& 100			& 0		&
M38	& 28					& 10$^{7}$		& 100			& 10$^{2.5}$	\\
M18	& 28					& 10$^{6}$		& 3000			& 0		&
M39	& 28					& 10$^{7}$		& 3000			& 10$^{2.5}$	\\
M19	& 160					& 10$^{7}$		& 1			& 0		&
M40	& 160					& 10$^{7}$		& 1			& 10$^{2.5}$	\\
M20	& 160					& 10$^{7}$		& 100			& 0		&
M41	& 160					& 10$^{7}$		& 100			& 10$^{2.5}$	\\
M21	& 160					& 10$^{7}$		& 3000			& 0		&
M42	& 160					& 10$^{7}$		& 3000			& 10$^{2.5}$	\\
\hline
\end{tabular}
    \label{tab:table1}
\end{center}
\end{table*}

\subsection{The initial conditions}
We created a simulation box of size 24$^{3}$ pc$^{3}$ with outflow boundaries in each direction and isolated in terms of gravity. In here, we put a typical molecular cloud, or clump as some prefer, with a uniform number density of $\rm 10^{5} \rm ~cm^{-3}$ and a size of $\rm r_{\rm cloud}=0.33$ pc in spherical radius. With a mean molecular weight of $\rm \mu$ = 2.3, the total mass of the cloud amounts to 800 solar masses. The rest of the medium is filled with gas that has a uniform density of 10 $\rm cm^{-3}$. The total gas mass of the simulation box is 8000 solar masses. At the center of the box, we put a point particle with mass $\rm M_{\rm bh}$ that represents the black hole, where $\rm M_{\rm bh}=[10^{6},10^{7},10^{8}] ~M_{\odot}$. The temperature of the gas is initialized as 10 K, but depending on the model we expose the cloud, immediately after the simulation starts, to external radiation. Since these radiative processes are fast, with respect to the hydrodynamics of the simulation, the temperature changes quickly, within $\rm 10^{8} \rm ~s$ (= one timestep), after initialization. The radiation source is either an X-ray emitter (accreting black hole), a uniform background of cosmic rays (mainly supernova remnants), and/or an isotropic UV radiation field (nearby massive stars).

An initially random, divergence-free turbulent velocity field is applied to the molecular cloud with a characteristic FWHM of 5 km/s that agrees well with clouds observed in active regions \citep{2009A&A...503..459P}. For the isothermal runs, the sound speed of the cloud is $\rm c_{\rm s}=0.19$ km/s (for T=10 K) and can go up to $\sim$5 km/s when radiation impinges on the cloud. There are supersonic flows with Mach numbers of up to 25. We apply the turbulence over all scales with a power spectrum of $\rm P(k) \propto k^{-4}$, as mentioned earlier, following the empirical laws for compressible fluids \citep{1981MNRAS.194..809L, 1999ApJ...522L.141M, 2004ApJ...615L..45H}. The turbulence in this work is not driven. Still, it can remain strong throughout a simulation due to gravitational instabilities or shear induced by the black hole, and does so for the larger $\rm M_{bh}$ runs.

The simulations start with a cloud at 10 pc distance from the black hole that is in a stable Keplerian orbit. The orbital time is on the order of $\rm 10^{6} \times (10^{7} ~M_{\odot}/M_{bh})^{1/2}$ yr, which is larger than the cloud free-fall time in any of the models. Shear due to the black hole gravity, creates a maximum velocity difference of about $\rm \triangle v_{\rm shear}=2.22 \times (M_{bh}/10^{7} ~M_{\odot})^{1/2}$ km/s across the cloud. This follows from

\begin{equation} \rm
\triangle v_{\rm shear}=\sqrt{ \rm GM_{\rm bh}r_{\rm cloud}^{2} \over d_{\rm bh}^{3} },
\end{equation}

\noindent
where $\rm r_{cloud} \ll d_{bh}$. The shearing time, $\rm t_{\rm shear}=2r_{\rm cloud}/\triangle \rm v_{\rm shear}$, is about 1-10 times larger than the cloud free-fall time and gravitationally bound (roughly) spherical clouds are thus likely to exist at densities of $\sim$10$^{5} \rm ~cm^{-3}$.

The maximum grid resolution that we allow for in any simulation is 4096$^{3}$ cells. For a box of size 24 pc, the maximum spatial resolution becomes $\rm \triangle x = 1.76 \times 10^{16}$ cm. All the simulations are evolved up to 3 free-fall times, where one free-fall time, given the initial conditions, is $\rm t_{\rm ff} = 10^{5}$ years. This is taken as the basic time unit throughout this work.

\section{Results}
\label{sec:results}
\begin{figure*}[htb!]
\hspace{-0.7cm}
\begin{tabular}{lll}

\vspace{0.7cm}

\begin{minipage}{6.0cm}
\includegraphics[scale=0.39]{phase-m01-x0m7u0c1chk-30.epsi}
\end{minipage} &
\begin{minipage}{6.0cm}
\includegraphics[scale=0.39]{phase-m02-x0m7u0c100chk-30.epsi}
\end{minipage} &
\begin{minipage}{6.0cm}
\includegraphics[scale=0.39]{phase-m03-x0m7u0c3000chk-30.epsi}
\end{minipage} \\

\vspace{0.7cm}

\begin{minipage}{6.0cm}
\includegraphics[scale=0.39]{phase-m07-x5.1m7u0c1chk-30.epsi}
\end{minipage} &
\begin{minipage}{6.0cm}
\includegraphics[scale=0.39]{phase-m08-x5.1m7u0c100chk-30.epsi}
\end{minipage} &
\begin{minipage}{6.0cm}
\includegraphics[scale=0.39]{phase-m09-x5.1m7u0c3000chk-32.epsi}
\end{minipage} \\

\vspace{0.7cm}

\begin{minipage}{6.0cm}
\includegraphics[scale=0.39]{phase-m21-x28m7u0c1chk-30.epsi}
\end{minipage} &
\begin{minipage}{6.0cm}
\includegraphics[scale=0.39]{phase-m22-x28m7u0c100chk-31.epsi}
\end{minipage} &
\begin{minipage}{6.0cm}
\includegraphics[scale=0.39]{phase-m23-x28m7u0c3000chk-30.epsi}
\end{minipage} \\

\begin{minipage}{6.0cm}
\includegraphics[scale=0.39]{phase-m13-x160m7u0c1chk-30.epsi}
\end{minipage} &
\begin{minipage}{6.0cm}
\includegraphics[scale=0.39]{phase-m14-x160m7u0c100chk-30.epsi}
\end{minipage} &
\begin{minipage}{6.0cm}
\includegraphics[scale=0.39]{phase-m15-x160m7u0c3000chk-30.epsi}
\end{minipage} \\

\begin{minipage}{6.0cm}
\includegraphics[scale=0.67]{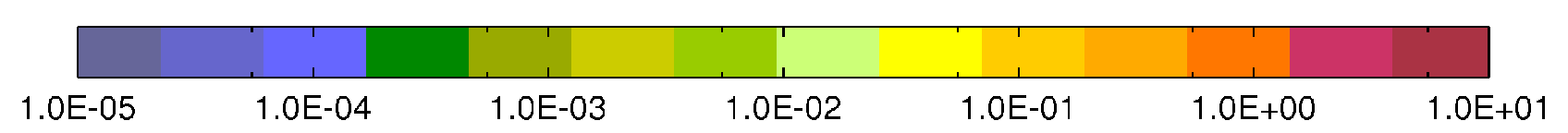}
\end{minipage} \\

\end{tabular}
\caption{Phase diagrams for the models with $\rm M_{\rm bh}=10^{7} ~M_{\odot}$ and UV=0. Temperature is plotted against number density at $\rm t=t_{ff}=10^{5} ~yr$. The diagrams are gridded into $\rm 75^{2}$ cells with the weighted masses of the points depicted in color. Red represents a mass of $\rm 10 ~M_{\odot}$ or above and blue is for $\rm 10^{-5} ~M_{\odot}$. Isothermal conditions yield a flat profile. From left to right, the cosmic ray rate increases from 1, 100 to 3000 $\rm \times ~Galactic$. From top to bottom, the X-ray flux increases from 0, 5.1, 28 to 160 $\rm erg ~s^{-1} ~cm^{-2}$.}
\label{fig:phases1}
\end{figure*}

\begin{figure*}[htb!]
\hspace{-0.7cm}
\begin{tabular}{lll}

\vspace{0.7cm}

\begin{minipage}{6.0cm}
\includegraphics[scale=0.39]{phase-m04-x0m6u0c1chk-30.epsi}
\end{minipage} &
\begin{minipage}{6.0cm}
\includegraphics[scale=0.39]{phase-m05-x0m6u0c100chk-30.epsi}
\end{minipage} &
\begin{minipage}{6.0cm}
\includegraphics[scale=0.39]{phase-m06-x0m6u0c3000chk-30.epsi}
\end{minipage} \\

\vspace{0.7cm}

\begin{minipage}{6.0cm}
\includegraphics[scale=0.39]{phase-m10-x5.1m6u0c1chk-31.epsi}
\end{minipage} &
\begin{minipage}{6.0cm}
\includegraphics[scale=0.39]{phase-m11-x5.1m6u0c100chk-30.epsi}
\end{minipage} &
\begin{minipage}{6.0cm}
\includegraphics[scale=0.39]{phase-m12-x5.1m6u0c3000chk-31.epsi}
\end{minipage} \\

\vspace{0.7cm}

\begin{minipage}{6.0cm}
\includegraphics[scale=0.39]{phase-m24-x28m6u0c1chk-32.epsi}
\end{minipage} &
\begin{minipage}{6.0cm}
\includegraphics[scale=0.39]{phase-m25-x28m6u0c100chk-30.epsi}
\end{minipage} &
\begin{minipage}{6.0cm}
\includegraphics[scale=0.39]{phase-m26-x28m6u0c3000chk-30.epsi}
\end{minipage} \\

\begin{minipage}{6.0cm}
\includegraphics[scale=0.39]{phase-m16-x160m6u0c1chk-30.epsi}
\end{minipage} &
\begin{minipage}{6.0cm}
\includegraphics[scale=0.39]{phase-m17-x160m6u0c100chk-28.epsi}
\end{minipage} &
\begin{minipage}{6.0cm}
\includegraphics[scale=0.39]{phase-m18-x160m6u0c3000chk-30.epsi}
\end{minipage} \\

\begin{minipage}{6.0cm}
\includegraphics[scale=0.67]{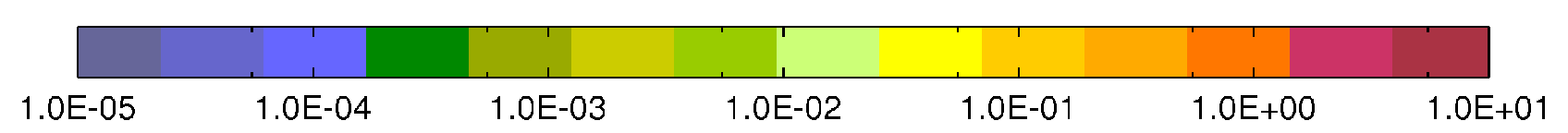}
\end{minipage} \\

\end{tabular}
\caption{Phase diagrams for the models with $\rm M_{\rm bh}=10^{6} ~M_{\odot}$ and UV=0. Temperature is plotted against number density at $\rm t=t_{ff}=10^{5} ~yr$. The diagrams are gridded into $\rm 75^{2}$ cells with the weighted masses of the points depicted in color. Red represents a mass of $\rm 10 ~M_{\odot}$ or above and blue is for $\rm 10^{-5} ~M_{\odot}$. Isothermal conditions yield a flat profile. From left to right, the cosmic ray rate increases from 1, 100 to 3000 $\rm \times ~Galactic$. From top to bottom, the X-ray flux increases from 0, 5.1, 28 to 160 $\rm erg ~s^{-1} ~cm^{-2}$.}
\label{fig:phases2}
\end{figure*}

\begin{figure*}[htb!]
\hspace{-0.7cm}
\begin{tabular}{lll}

\vspace{0.7cm}

\begin{minipage}{6.0cm}
\includegraphics[scale=0.39]{phase-m31-x0m8u0c1chk-30.epsi}
\end{minipage} &
\begin{minipage}{6.0cm}
\includegraphics[scale=0.39]{phase-m32-x0m8u0c100chk-30.epsi}
\end{minipage} &
\begin{minipage}{6.0cm}
\includegraphics[scale=0.39]{phase-m33-x0m8u0c3000chk-30.epsi}
\end{minipage} \\

\begin{minipage}{6.0cm}
\includegraphics[scale=0.39]{phase-m40-x160m8u0c1chk-31.epsi}
\end{minipage} &
\begin{minipage}{6.0cm}
\includegraphics[scale=0.39]{phase-m41-x160m8u0c100chk-31.epsi}
\end{minipage} &
\begin{minipage}{6.0cm}
\includegraphics[scale=0.39]{phase-m42-x160m8u0c3000chk-30.epsi}
\end{minipage} \\

\begin{minipage}{6.0cm}
\includegraphics[scale=0.67]{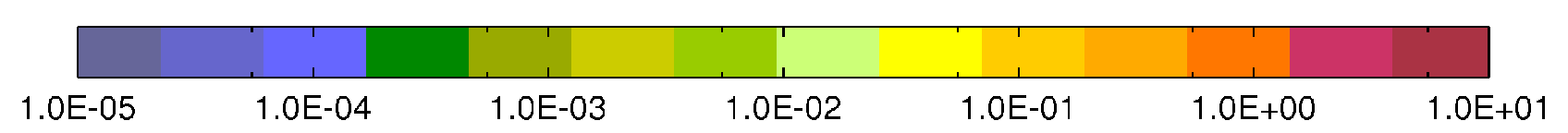}
\end{minipage} \\

\end{tabular}
\caption{Phase diagrams for the models with $\rm M_{\rm bh}=10^{8} ~M_{\odot}$ and UV=0. Temperature is plotted against number density at $\rm t=t_{ff}=10^{5} ~yr$. The diagrams are gridded into $\rm 75^{2}$ cells with the weighted masses of the points depicted in color. Red represents a mass of $\rm 10 ~M_{\odot}$ or above and blue is for $\rm 10^{-5} ~M_{\odot}$. Isothermal conditions yield a flat profile. From left to right, the cosmic ray rate increases from 1, 100 to 3000 $\rm \times ~Galactic$. From top to bottom, the X-ray flux increases from 0 to 160 $\rm erg ~s^{-1} ~cm^{-2}$.}
\label{fig:phases3}
\end{figure*}

\begin{figure*}[htb!]
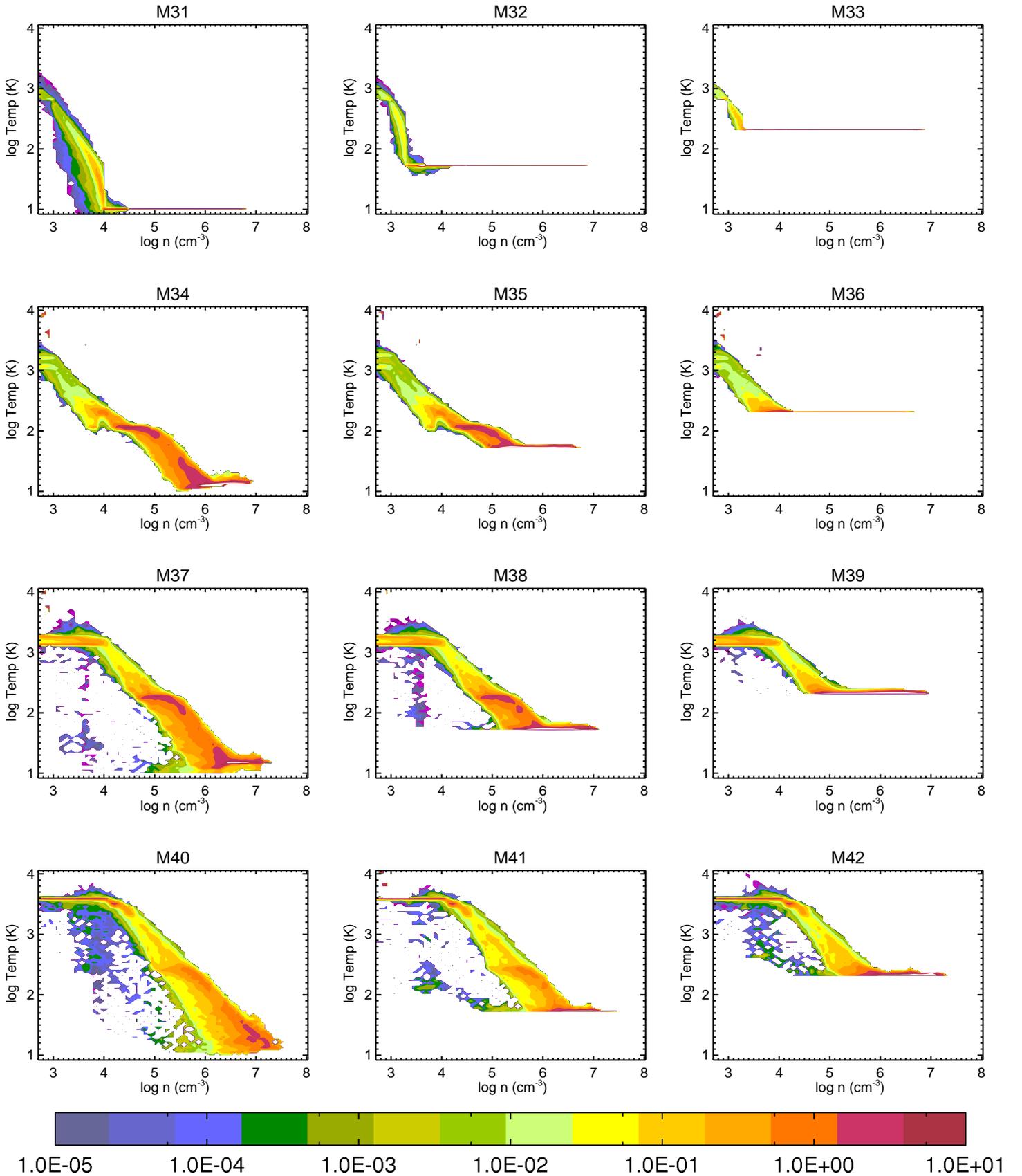

\hspace{-0.7cm}
\begin{tabular}{lll}

\vspace{0.7cm}

\begin{minipage}{6.0cm}
\includegraphics[scale=0.39]{phase-m51-x0m7u2.5c1chk-30.epsi}
\end{minipage} &
\begin{minipage}{6.0cm}
\includegraphics[scale=0.39]{phase-m55-x0m7u2.5c100chk-30.epsi}
\end{minipage} &
\begin{minipage}{6.0cm}
\includegraphics[scale=0.39]{phase-m59-x0m7u2.5c3000chk-30.epsi}
\end{minipage} \\

\vspace{0.7cm}

\begin{minipage}{6.0cm}
\includegraphics[scale=0.39]{phase-m52-x5.1m7u2.5c1chk-30.epsi}
\end{minipage} &
\begin{minipage}{6.0cm}
\includegraphics[scale=0.39]{phase-m56-x5.1m7u2.5c100chk-28.epsi}
\end{minipage} &
\begin{minipage}{6.0cm}
\includegraphics[scale=0.39]{phase-m60-x5.1m7u2.5c3000chk-30.epsi}
\end{minipage} \\

\vspace{0.7cm}

\begin{minipage}{6.0cm}
\includegraphics[scale=0.39]{phase-m53-x28m7u2.5c1chk-30.epsi}
\end{minipage} &
\begin{minipage}{6.0cm}
\includegraphics[scale=0.39]{phase-m57-x28m7u2.5c100chk-30.epsi}
\end{minipage} &
\begin{minipage}{6.0cm}
\includegraphics[scale=0.39]{phase-m61-x28m7u2.5c3000chk-31.epsi}
\end{minipage} \\

\begin{minipage}{6.0cm}
\includegraphics[scale=0.39]{phase-m54-x160m7u2.5c1chk-30.epsi}
\end{minipage} &
\begin{minipage}{6.0cm}
\includegraphics[scale=0.39]{phase-m58-x160m7u2.5c100chk-30.epsi}
\end{minipage} &
\begin{minipage}{6.0cm}
\includegraphics[scale=0.39]{phase-m62-x160m7u2.5c3000chk-30.epsi}
\end{minipage} \\

\begin{minipage}{6.0cm}
\includegraphics[scale=0.67]{colorbar-38.ps}
\end{minipage} \\

\end{tabular}
\caption{Phase diagrams for the models including UV. The UV flux used in these models is $\rm 10^{2.5} ~G_{0}$. Similar to the figures before, Figs. \ref{fig:phases1} to \ref{fig:phases3}, the images display the temperature against number density at $\rm t=t_{ff}=10^{5} ~yr$. The diagrams are gridded into $\rm 75^{2}$ cells with the weighted masses of the points depicted in color. Red represents a mass of $\rm 10 ~M_{\odot}$ or above and blue is for $\rm 10^{-5} ~M_{\odot}$. From left to right, the cosmic ray rate increases from 1, 100 to 3000 $\rm \times ~Galactic$. From top to bottom, the X-ray flux increases from 0 to 160 $\rm erg ~s^{-1} ~cm^{-2}$. These UV models are simulated for a black hole mass of $\rm M_{\rm bh}=10^{7} ~M_{\odot}$.}
\label{fig:phases-uv}
\end{figure*}

We divide the results into four sections. In each section we individually evaluate the effects of X-rays, cosmic rays, UV, and shear, while keeping every other parameter fixed. In Figs. \ref{fig:phases1} to \ref{fig:imfs-uv}, we first show the results of all parameter variations at once, as this gives a better overview. Figs. \ref{fig:phases1} to \ref{fig:phases-uv} display the temperature-density phase diagrams, Figs. \ref{fig:sfes1} to \ref{fig:sfes-uv} the star-formation efficiencies, and Figs. \ref{fig:imfs1} to \ref{fig:imfs-uv} the initial mass functions.

For each phase diagram in Figs. \ref{fig:phases1} to \ref{fig:phases-uv} we plot the number density versus the temperature for one moment in time, which is at $\rm 1 ~t_{ff}$. The diagram is subsequently gridded into $\rm 75^{2}$ cells and the weighted masses of all the points within each cell is summed up and depicted in color. Note that the isothermal conditions always yield a flat profile.

The star-formation efficiencies in Figs. \ref{fig:sfes1} to \ref{fig:sfes-uv} are displayed by plotting the ratio of the total sink particle mass over the total initial gas mass (8000 $\rm M_{\odot}$), $\rm {\rm SFE}=M_{\rm sink}/M_{\rm total}$, against time. The efficiencies are plotted as they are, i.e., no fitting is involved. Generally between 1 and 2 $\rm t_{\rm ff}$, the star-formation efficiencies flatten out. This is due to the depletion of high density gas. At this stage, star formation is almost completely halted and accretion is reduced to a minimum.

Each IMF plot in Figs. \ref{fig:imfs1} to \ref{fig:imfs-uv} is constructed by taking all the sink-particles of 61 different cloud evolution times of one model and by normalizing them. For this, we took the time frames between 1 $\rm t_{\rm ff}$ and 3 $\rm t_{\rm ff}$, with a time resolution of 1/30 $\rm t_{\rm ff}$. The sink particle masses are then plotted as a logarithmically binned histogram, with a fixed number of 16 bins, of stellar mass versus number. We compared these IMFs with those corresponding to the time frames between 1 $\rm t_{\rm ff}$ and 2 $\rm t_{\rm ff}$, and found that there was little difference between them. There are 4 colored lines plotted in each of the sub-figures. The green line represents the Salpeter IMF with a slope of $\rm \Gamma_{\rm Sal}=-1.35$ and is shown for comparison purposes only. The blue line displays the Chabrier IMF as fitted to our fiducial case, model $\rm M01$, an isothermal simulation at $\rm t=t_{\rm ff}$ (this is a single snapshot in time) with $\rm M_{bh}=10^{7} ~M_{\odot}, F_{X}=0$, $\rm F_{UV}=0$, and $\rm \zeta_{\rm CR}=1\times \rm Galactic$. See also \cite{2010A&A...522A..24H}. This curve is a lognormal function up to 1 $\rm M_{\odot}$ with a power-law tail of $\rm \Gamma_{\rm Cha}=-1.30$ \citep{2003PASP..115..763C}. The red solid line gives the best (least squares) lognormal fit, whereas the purple triple-dotted line shows the best (least squares) power-law fit above the turn-over mass. The turn-over mass is not fixed and can change for each model. We determine the turn-over mass by finding the position with the lowest absolute derivative (slope) $\rm |dlogN/dlogM| \rightarrow 0$ from the lognormal fit.

\subsection{Effects of X-rays}
X-rays heat up the lower density unshielded gas and can do so up to 6000 K at the highest flux. Hard X-rays ($>$1 keV) have a high penetration factor. They can pierce through columns of gas of up to $\rm N\simeq10^{24} ~cm^{-2}$ \citep{2005A&A...436..397M}. Since the X-ray source is in the center of the simulation box, material that lies behind high density gas, with column densities beyond $\rm 10^{23} \rm ~cm^{-2}$, can be strongly shielded from the radiation. This actually occurs in our simulations. Our initialized model cloud has a maximum column of $\rm N = 2.0\cdot10^{23} \rm ~cm^{-2}$ ($\rho = 10^{5} \rm ~cm^{-3}$ and $\rm 2r_{\rm cloud} = 2.0\cdot10^{18} \rm ~cm$). Some of the phase diagrams show a secondary line at low densities that exhibits a sharp drop to low temperatures because of the cold shielded gas. In each of the Figs. \ref{fig:phases1} to \ref{fig:imfs-uv} the X-rays increase from top to bottom.

From the phase diagrams, it is immediately clear that the gas temperatures decrease with increasing density and that the temperatures are higher for the higher X-ray fluxes at any point in density. The latter is a direct consequence of the efficient Coulomb heating by X-rays. X-rays do not only heat the system, but also allow new paths for cooling to proceed along. X-rays increase the ionization fraction of the species and new molecules are formed. This results in a higher cooling efficiency with increasing density, hence, the decreasing temperature trend. We can also see a spread in the temperature as a direct effect from the different column densities throughout the cloud. The spread seems to be larger for the higher X-ray fluxes, however, this is not resulting from a larger column density range, but is merely due to the larger range in temperatures from the increased X-ray flux. An interesting feature is that there is a range in densities around $\rm n \sim 10^{5} ~cm^{-3}$ where the temperature lingers around a few 100 K. The cooling process is slowed down here due to LTE effects (thermalization and line trapping) until the densities are high enough so that cooling can start to be effective again through gas-dust coupling.

\begin{figure}[htb!]
\centering
\includegraphics[scale=0.52]{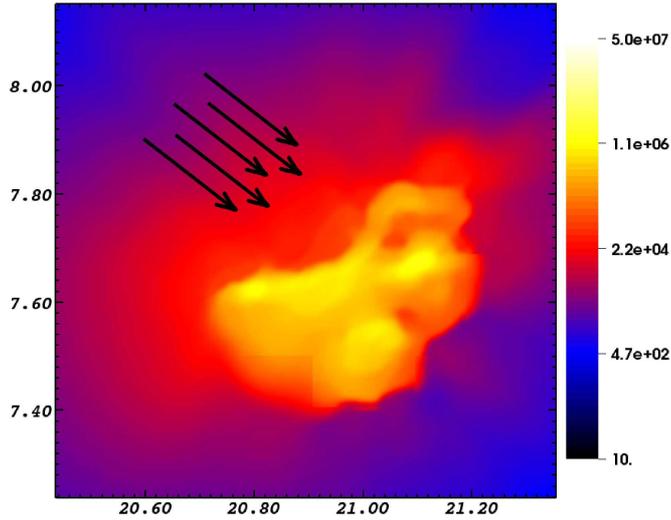}
\caption{Density morphology of model M19, i.e., with $\rm F_{X}=160 \rm ~erg ~s^{-1} ~cm^{-2}$, at t=2/3t$_{\rm ff}$. The image shows a slice through the XY-plane of the center of the cloud. The color represents the number density (cm$^{-3}$) and the axes are in parsec. The arrows represent the direction of radiation emanating from the black hole, which is located at the upper left side.}
\label{fig:compressed}
\end{figure}

From the color in the phase diagrams we can infer that most of the gas mass lies at high densities, but for higher X-ray fluxes a lot of mass also lies in the low density, high temperature regime. This is the area where the cloud is directly irradiated by X-rays and the cloud evaporates. As a consequence, the cloud size and its mass are reduced by this. This has repercussions for the final star-formation efficiency. When comparing an isothermal model against the model with the highest X-ray flux, for $\rm \zeta_{CR}=1$, we see that the SFE is reduced from 9.4\% to 8.2\% for a $\rm 10^{7} ~M_{\odot}$ black hole at $\rm t=3t_{\rm ff}$, and from 12.3\% to 7.0\% for a $\rm 10^{6} ~M_{\odot}$ black hole (Figs. \ref{fig:sfes1} and \ref{fig:sfes2}). The same behavior is found for higher cosmic ray rates (discussed in section \ref{sec:cosmic rays}). Despite some fluctuations due to the randomness of the system, it is interesting to see that the efficiency generally increases a little with our lowest X-ray flux compared to the isothermal models. It seems that a mild X-ray flux of 5.1 $\rm erg ~s^{-1} ~cm^{-2}$ or less actually enhances the star-formation efficiency. See Table \ref{tab:table2} for the list of results. Although gas temperatures are higher and the SFE decreases when going to higher X-ray fluxes, the star-formation rate, $\rm {\rm SFR} = d{\rm SFE}/dt$, is still high early on. We can see from the SFE plots that star formation is initiated at about the same time, $\rm t_{onset}\simeq0.6t_{ff}$, irrespective of X-rays. We find that the quenching of star formation due to X-ray heating is balanced by the increased densities resulting from the thermal pressure that the radiation field creates. An ionizing pressure front compresses the cloud which leads to very efficient star formation, similar to, if not more than, the colder, isothermal, cloud models. Fig. \ref{fig:compressed} shows a 2D slice to this effect.

We see that the IMFs tend to get flatter for the higher X-ray fluxes. The best power-law fit gives us a nice Salpeter slope with $\rm \Gamma_{powfit}=-1.34$ for model M01, i.e., without X-rays, and a slope of $\rm \Gamma_{powfit}=-1.31,-1.24,-1.01$ for models M07, M13, and M19, i.e., for fixed parameters but with increasing X-ray flux. A lognormal distribution also seems to fit most of the IMFs quite well. The same trend is visible for all other cloud conditions. This implies that the IMF shape depends on how much the cloud is irradiated. Another effect is that the turn-over mass shifts toward higher masses with increasing flux, from $\rm M_{turn} = 0.54 ~M_{\odot}$, at the lowest X-ray flux (M07), to about 0.70 and 1.51 $\rm M_{\odot}$ for the higher fluxes (M13 and M19). However, the corresponding isothermal fiducial model (M01) seems to have a bit higher turn-over mass, 0.66 $\rm M_{\odot}$, than our lowest X-ray flux (5.1 $\rm erg ~s^{-1} ~cm^{-2}$) model. Again, a similar trend is seen for all cloud models.

\begin{table*}[htb]
\begin{center}
\caption{List of results for each model}
\begin{tabular}{ccccccc}
\hline
\hline
Model	& IMF slope & Turn-over mass & Onset of SF & Final SFE & Accreted fraction & Number of particles \\
& & [$\rm M_{\odot}$]	& [$\rm t_{\rm ff}=10^{5} \rm ~yr$]	& [$\rm M_{\rm part.}/M_{\rm tot.}$]	& 	& \\
\hline
M01	& -1.34			& 0.66		& 0.47 	& 0.094 	& 0.606		& 247 \\
M02	& -1.44			& 2.37		& 0.87 	& 0.092 	& 0.691		& 110 \\
M03	& ---$^{a}$		& ~4.25$^{a}$	& 1.16 	& 0.089 	& 0.747		& 27 \\
M04	& -1.30			& 0.47		& 0.35 	& 0.123 	& 0.801		& 262 \\
M05	& -1.24			& 2.60		& 0.67 	& 0.125 	& 0.795		& 63 \\
M06	& ---$^{a}$		& ~5.95$^{a}$	& 0.99 	& 0.117 	& 0.725		& 21 \\
M07	& -1.31			& 0.54		& 0.43 	& 0.099 	& 0.609		& 350 \\
M08	& -1.08			& 0.65		& 0.50 	& 0.095 	& 0.693		& 227 \\
M09	& -1.07			& 0.65		& 0.60 	& 0.077 	& 0.653		& 211 \\
M10	& -1.04			& 0.43		& 0.36 	& 0.125 	& 0.768		& 276 \\
M11	& -1.05			& 0.52		& 0.46 	& 0.117 	& 0.708		& 137 \\
M12	& -0.78			& 0.26		& 0.52 	& 0.116 	& 0.762		& 104 \\
M13	& -1.24			& 0.70		& 0.52 	& 0.102 	& 0.607		& 286 \\
M14	& -1.02			& 0.73		& 0.59 	& 0.092 	& 0.597		& 211 \\
M15	& -0.92			& 0.77		& 0.67 	& 0.093 	& 0.619		& 171 \\
M16	& -1.16			& 0.68		& 0.42 	& 0.104 	& 0.598		& 202 \\
M17	& -1.01			& 0.65		& 0.52 	& 0.100 	& 0.545		& 153 \\
M18	& -0.58			& 0.52		& 0.59 	& 0.094 	& 0.592		& 104 \\
M19	& -1.01			& 1.51		& 0.65 	& 0.082 	& 0.637		& 138 \\
M20	& -0.91			& 1.62		& 0.65 	& 0.076 	& 0.620		& 134 \\
M21	& -0.85			& 1.43		& 0.67 	& 0.073 	& 0.621		& 111 \\
M22	& -0.82			& 1.12		& 0.55 	& 0.070 	& 0.647		& 80 \\
M23	& -0.70			& 1.91		& 0.57 	& 0.071 	& 0.623		& 79 \\
M24	& -0.62			& 1.36		& 0.62 	& 0.073 	& 0.627		& 58 \\
M25	& -1.32			& 1.20		& 0.74	& 0.014		& 0.476		& 84 \\
M26	& ---$^{a}$		& ---$^{a}$	& 0.80	& 0.001		& 0.681		& 3 \\
M27	& ---$^{b}$		& ---$^{b}$	& ---$^{b}$& ---$^{b}$	& ---$^{b}$	& 0 \\
M28	& -1.30			& 1.62		& 0.66	& 0.037		& 0.637		& 108 \\
M29	& -1.28			& 1.86		& 0.66	& 0.024		& 0.697		& 59 \\
M30	& -1.07			& 1.53		& 0.65 	& 0.013		& 0.618		& 49 \\
M31	& -1.48			& 0.55		& 0.49	& 0.103		& 0.638		& 262 \\
M32	& -1.31			& 2.30		& 0.90	& 0.099		& 0.728		& 106 \\
M33	& ---$^{a}$		& 4.30$^{a}$	& 1.16	& 0.093		& 0.850		& 19 \\
M34	& -1.23			& 0.43		& 0.40	& 0.101		& 0.568		& 332 \\
M35	& -0.99			& 0.57		& 0.54	& 0.085		& 0.658		& 230 \\
M36	& -1.06			& 0.65		& 0.60	& 0.096		& 0.638		& 198 \\
M37	& -1.05			& 0.59		& 0.51	& 0.107		& 0.650		& 281 \\
M38	& -1.04			& 0.66		& 0.56	& 0.101		& 0.647		& 207 \\
M39	& -0.99			& 0.88		& 0.67	& 0.101		& 0.748		& 163 \\
M40	& -1.40			& 1.58		& 0.66	& 0.084		& 0.641		& 142 \\
M41	& -1.06			& 1.47		& 0.66	& 0.082		& 0.648		& 132 \\
M42	& -0.79			& 1.52		& 0.67	& 0.069		& 0.614		& 103 \\
\hline
\end{tabular}
    \label{tab:table2}
\end{center}
\vspace{-0.3cm}
\hspace{1.7cm} $^{a}$ Too few particles have formed here to properly construct an IMF.

\hspace{1.7cm} $^{b}$ No particles have formed in this simulation.
\end{table*}

\subsection{Effects of cosmic rays}
\label{sec:cosmic rays}
Cosmic rays can easily penetrate a molecular cloud and heat the gas uniformally. As such, they set the minimum attainable temperature in these systems \citep{1978ApJ...222..881G, 2007ARA&A..45..339B}. For our model cloud conditions, which have cosmic ray rates of 1, 100, and 3000$\times$Galactic ($\zeta_{\rm CR,Gal} = 5\times10^{-17} \rm ~s^{-1}$), the minimum temperatures are 10 K, 50 K, and 200 K \citep{2011MNRAS.414.1705P, 2011A&A...525A.119M}. In each of the Figs. \ref{fig:phases1} to \ref{fig:imfs-uv} the cosmic ray rate increases from left to right.

\begin{figure}[htb!]
\centering
\includegraphics[scale=0.53]{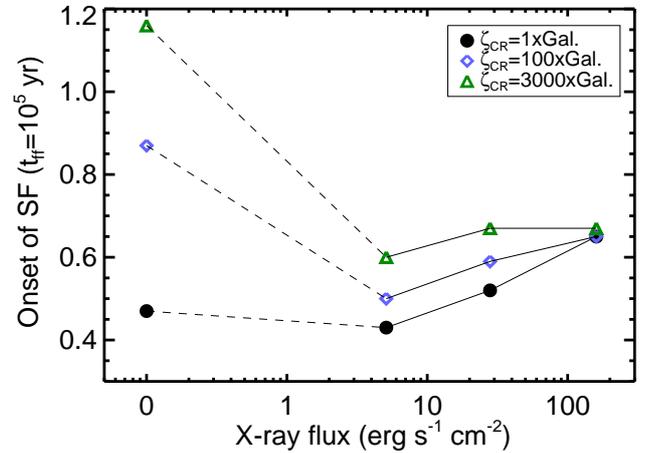}
\caption{Onset of star formation for the runs with a $\rm 10^{7} ~M_{\odot}$ black hole. In this figure, the star-formation initiation time is plotted against X-ray flux for three different cosmic ray rates. Higher cosmic ray rates delay the onset of star formation, while increasing the X-ray flux counteracts this. Note that the points on the left side are for the isothermal models which have an X-ray flux of 0.}
\label{fig:multiplot1}
\end{figure}

In contrast to the X-ray heating, cosmic rays clearly delay the onset of star formation. The first sink particles are formed much later with increasing $\rm \zeta_{\rm CR}$, see Fig. \ref{fig:multiplot1}. The number of sink particles formed is also drastically reduced by up to one order of magnitude, see Table \ref{tab:table2}. Despite this, once the cloud starts forming stars, it can do so rapidly and massively. The much higher mass each star obtains compensates for the loss in the number of stars. In the end, the SFEs between the different $\rm \zeta_{CR}$ runs are not much different. This can be explained by the fact that the Jeans mass scales strongly with the temperature and that the cloud fragments into star-forming cores that are more massive when the ambient temperature is higher. The stars that form out of them are therefore also heavier. These massive stars accrete more matter, since there is less competition around, and they tend to grow more. This result is also in agreement with the idea about fragmentation-induced starvation \citep{2010ApJ...725..134P, 2010arXiv1007.3530K}. But, we have a lack of fragmentation in this case and hence very little starvation.

In the absence of X-rays, one obvious result is that high cosmic ray rates strongly inhibit the formation of sub-solar mass sink particles, which follows from the aforementioned Jeans mass argument. The smallest mass we obtain in these runs is about 1 solar mass. This is in agreement with, and predicted by, \cite{2011MNRAS.414.1705P}. From the IMF plots, we see that the effect of cosmic rays on the slope of the mass function is minimal. The lognormal shape is also not much affected, although, given the relative low number of points at higher $\rm \zeta_{CR}$ values, one cannot firmly conclude this. It is thus so that there is a general shift in the mass distribution toward higher masses. The turn-over mass scales toward higher masses as well and has shifted from 0.66 $\rm M_{\odot}$ (M01) to 2.37 $\rm M_{\odot}$ (M02) for the $\rm M_{bh}=10^{7} ~M_{\odot}$ runs, and from 0.47 $\rm M_{\odot}$ (M04) to 2.60 $\rm M_{\odot}$ (M05) for the $\rm M_{bh}=10^{6} ~M_{\odot}$ runs. This is an increase of a factor of 3.6 and 5.5, respectively. Between these runs, the Jeans mass increases by a factor of $5^{3/2}=11.2$ (10 K to 50 K) at the same densities. So, for $\rm \zeta_{CR}=1$ to 100 times Galactic, it seems that the turn-over mass scales with roughly the temperature rather than the Jeans mass. We find that very massive sink particles are formed in the highest cosmic ray simulations. The most massive sink particle that formed in model M03 has a mass of 350 solar masses.

\subsection{Effects of UV}
In PDRs, UV radiation heats the gas up to a few thousand K and will dominate the chemistry of cloud surfaces. UV radiation can photodissociate molecules, which will lead to the destruction of efficient coolants, and heats through photo-electric emission from (small) dust grains. UV radiation is strongly attenuated above columns of $\rm N=10^{22} \rm ~cm^{-2}$, at solar metallicity. In our simulations, the model cloud has a starting column of $\rm N=10^{23} \rm ~cm^{-2}$ (center to edge). This grows to $\rm N=1\times10^{24} \rm ~cm^{-2}$, and beyond, when collapse is initiated and the first stars are about to form. Hence, it is not expected that UV radiation will have a direct effect on the results. Only by heating the low density regions, $\rm n < 10^{4} ~cm^{-3}$, and by pressurizing the medium where the molecular cloud is embedded in, the results might be influenced. We assume that there is a uniform background UV field. The results for the model clouds in UV dominated environments are shown in Figs. \ref{fig:phases-uv}, \ref{fig:sfes-uv}, and \ref{fig:imfs-uv}.

\begin{figure*}[htb!]
\hspace{-0.5cm}
\begin{tabular}{cc}
\includegraphics[scale=0.5]{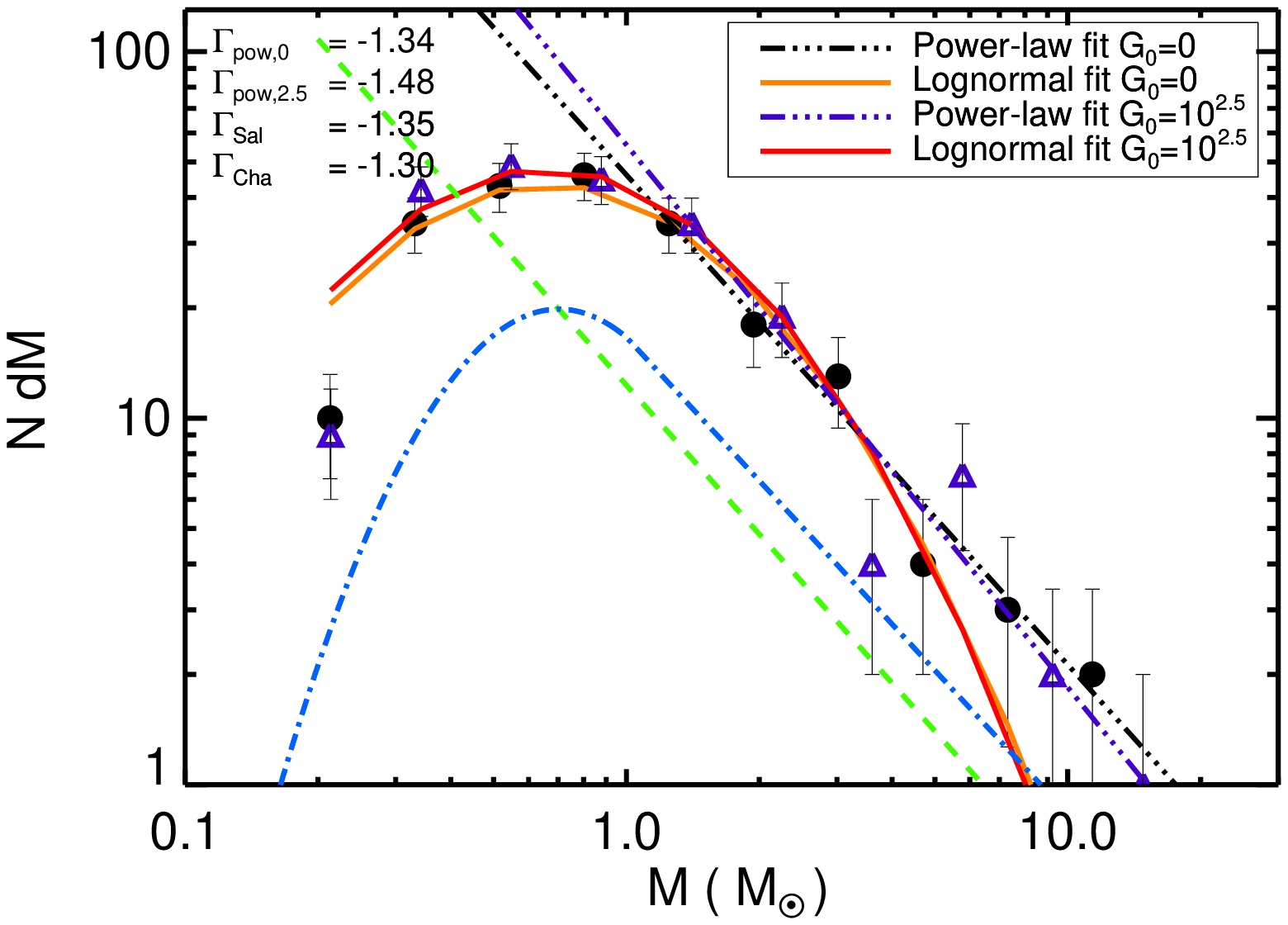}
\includegraphics[scale=0.5]{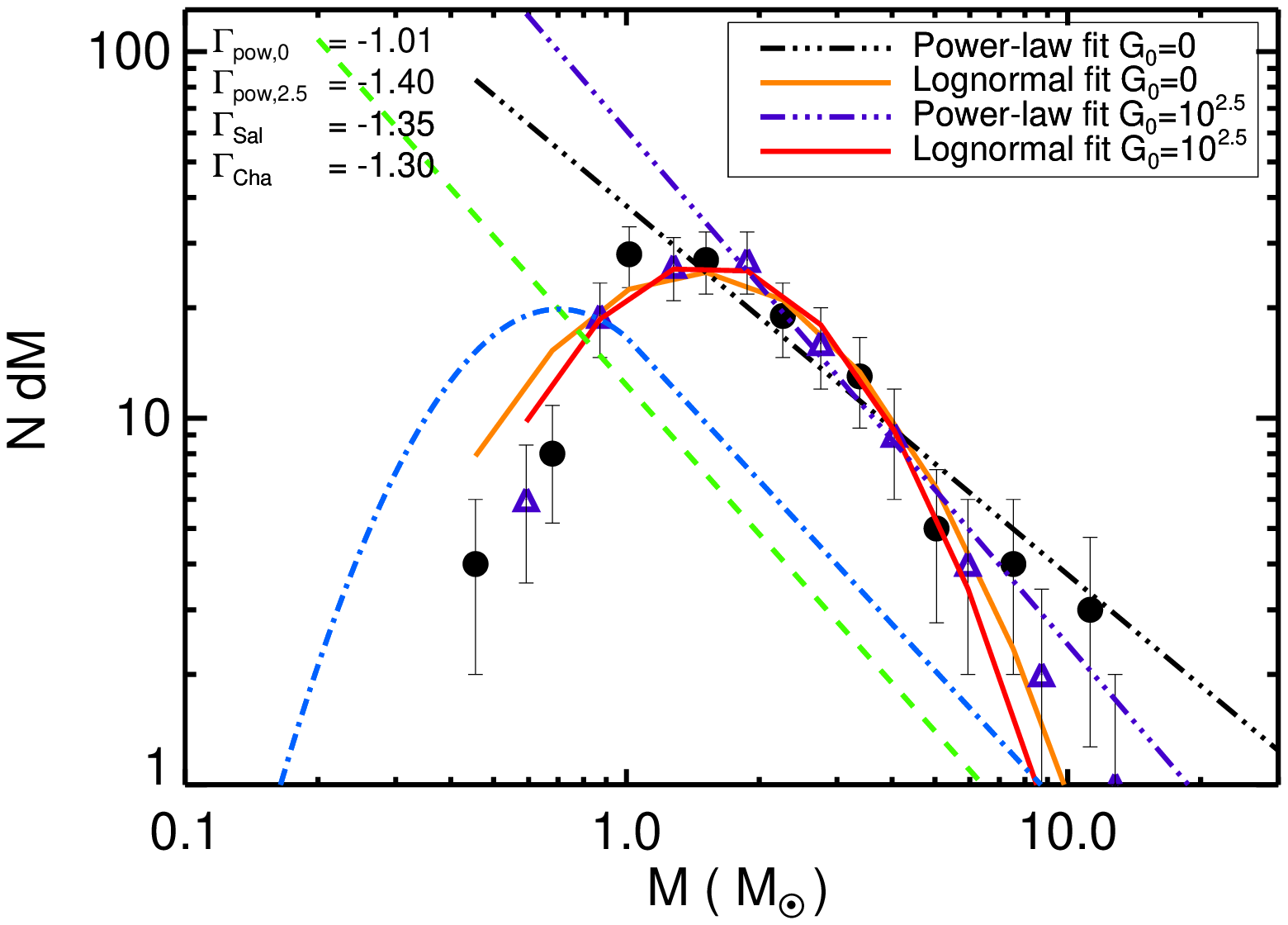}
\end{tabular}
\caption{Comparison of IMFs of the UV runs against their non-UV counterparts. The left panel shows the IMFs of the models M01 and M31 (no X-rays). The right panel shows the IMFs of the models M19 and M40 (highest X-ray flux). For comparison purposes, the Salpeter IMF (green dashed line) and the Chabrier IMF (blue dot-dashed line) are shown, as fitted to our fiducial case.}
\label{fig:uvimfsfe}
\end{figure*}

We have chosen to simulate the models with increased UV radiation for a single black hole mass, $\rm M_{bh}=10^{7}~M_{\odot}$, as can be seen from the table (Table \ref{tab:table1}). We have done so because our initial studies immediately showed that the direct effect of an external UV radiation field on our model clouds is modest. Follow-up simulations confirmed that the column density ($>$10$^{22} \rm ~cm^{-2}$) of the molecular cloud is, as predicted, already too high for UV radiation to penetrate even the edge of the cloud ($<$0.01 pc) and that turbulent effects and other instabilities do not provide the opportunity for UV radiation to do much damage. When X-rays are present, the gas heating is dominated by them. This is a consequence of the much higher (10-50\%) ionization heating efficiency compared to photo-electric heating (0.1-1\%) by UV irradiated dust grains. However, UV radiation does heat the low density gas, therefore, it influences the late time accretion.

The phase diagrams for the UV runs confirm that the radiation does not penetrate the natal cloud beyond a few magnitudes of visual extinction at one free-fall time. At low densities, we see that UV heats the gas to several hundred and upto a few thousand K. Above $\rm 10^{4} \rm ~cm^{-3}$ it has no impact. In the models together with X-rays, even with the lowest X-ray flux $\rm F_{X}=5.1 ~erg ~s^{-1} ~cm^{-2}$, there is little difference in the phase diagrams, as compared to the non-UV runs. Only the low density $\rm \leq 10^{4} ~cm^{-3}$, unshielded (in the UV) regions are heated up by UV. We also see that the SFE plots are not much different than their UV deficient counterparts. The slope of the IMF, on the other hand, is more sensitive to variations. We see that massive stars accrete less matter during the late stages of the run, i.e., $\rm t\gtrsim 2t_{ff}$, as the temperature is higher at low densities. This while most of the late time accretion normally comes from the low density, cold regions. Especially for the highest $\rm F_{X}$ runs, the slope of the IMF becomes slightly more steep due to a decrease in the mass of the most massive stars, in the presence of a UV background. In Fig. \ref{fig:uvimfsfe}, we plot the IMFs for four model runs, M01 versus M31 and M19 versus M40, that is, with and without UV, together in the same figure to highlight this effect.

\subsection{Effects of shear}
In the presence of strong gravity, the gravitational pull on each side of a bound extended object will cause a differential acceleration. If the kinetic energy resulting from the velocity difference is close to or on the order of the gravitational binding energy of the object, it will play an important role in the dynamics of the cloud. The object will be torn apart if the corresponding kinetic energy is much higher than its binding energy.

Our model cloud has a binding energy of

\begin{equation} \rm 
E_{\rm bind} = \frac{3GM_{\rm cloud}^{2}} {5r_{\rm cloud}} = \rm 1.02\times10^{47} ~erg,
\end{equation}

\noindent
where $\rm M_{\rm cloud}$ and $\rm r_{\rm cloud}$ are the cloud mass and radius. The maximum shear for the models with the three different black hole masses, for a cloud that is at 10 pc distance, results in kinetic energies of $\rm E_{\rm shear}=\rm [4\times10^{45},4\times10^{46},4\times10^{47}] ~erg$, with $\rm \triangle \rm v_{\rm shear}=[0.7, 2.2, 7.0]$ km/s. The velocity difference resulting from the gravitational stresses of the two higher black hole masses, $\rm M_{\rm bh}=[10^{7},10^{8}] ~M_{\odot}$, is on the order of the applied initial turbulence (5 km/s), and will therefore play an important role in the evolution of the cloud. Whereas the velocity shear due to a black hole with $\rm M_{\rm bh}=10^{6} ~M_{\odot}$ will have very little impact on the dynamics of the simulation. Figs. \ref{fig:phases1}, \ref{fig:sfes1}, and \ref{fig:imfs1} display the models with $\rm M_{\rm bh}=10^{7} ~M_{\odot}$. Figs. \ref{fig:phases2}, \ref{fig:sfes2}, and \ref{fig:imfs2} display the models with $\rm M_{\rm bh}=10^{6} ~M_{\odot}$. While, Figs. \ref{fig:phases3}, \ref{fig:sfes3}, and \ref{fig:imfs3} show the models with $\rm M_{\rm bh}=10^{8} ~M_{\odot}$.

\begin{figure*}[htb!]
\centering
\begin{tabular}{ccc}
\includegraphics[scale=0.25]{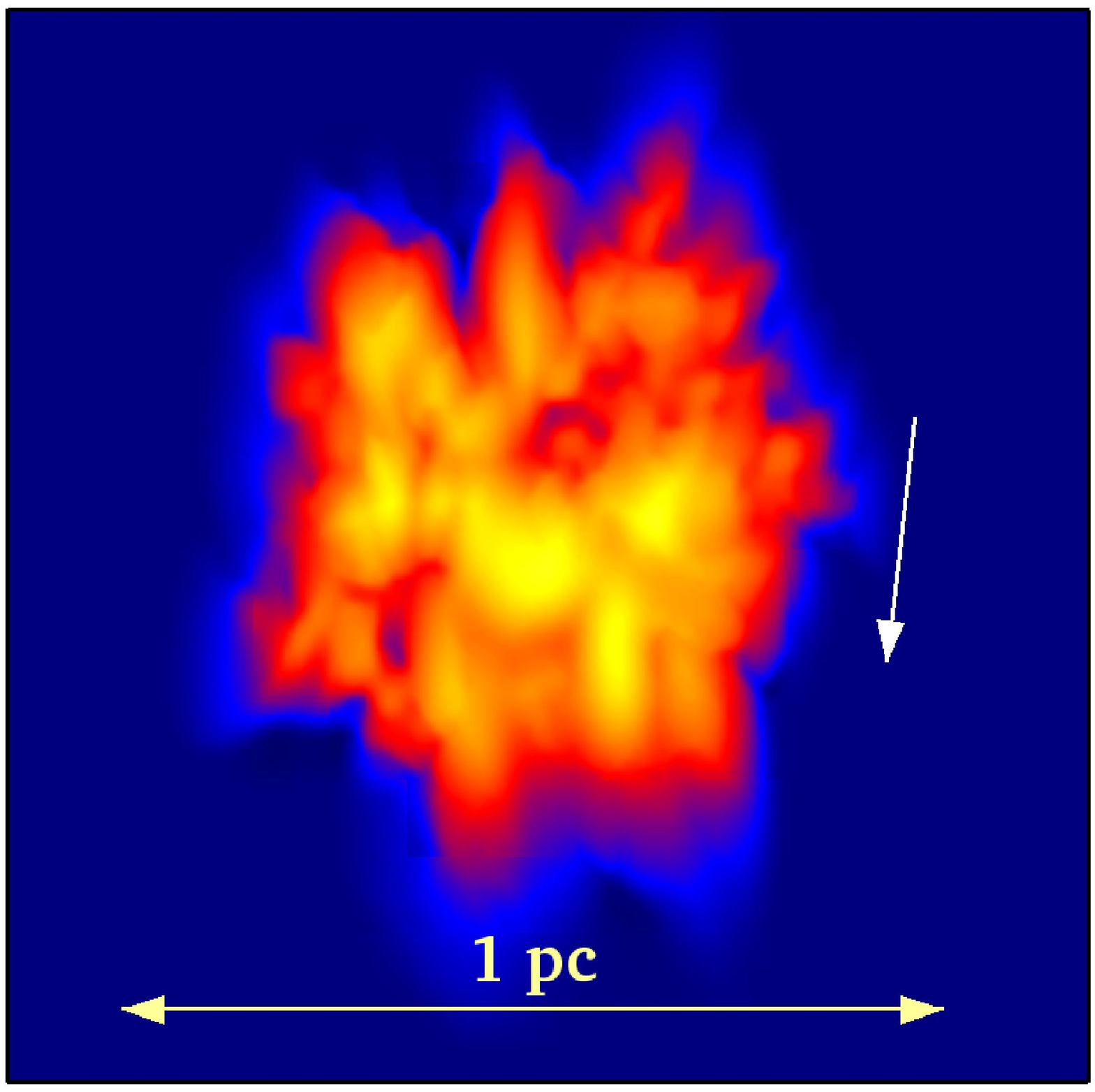}
\includegraphics[scale=0.25]{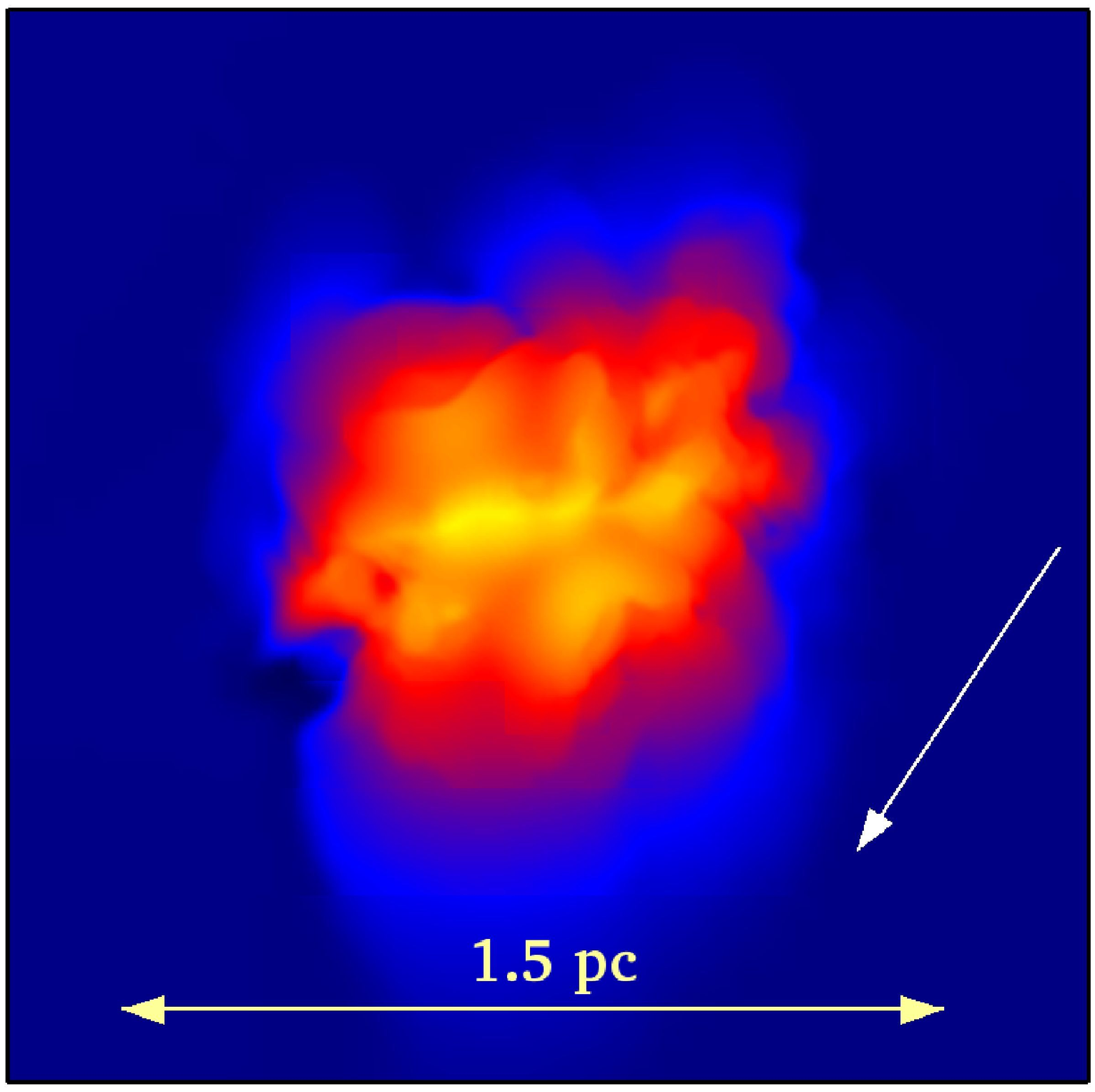}
\includegraphics[scale=0.25]{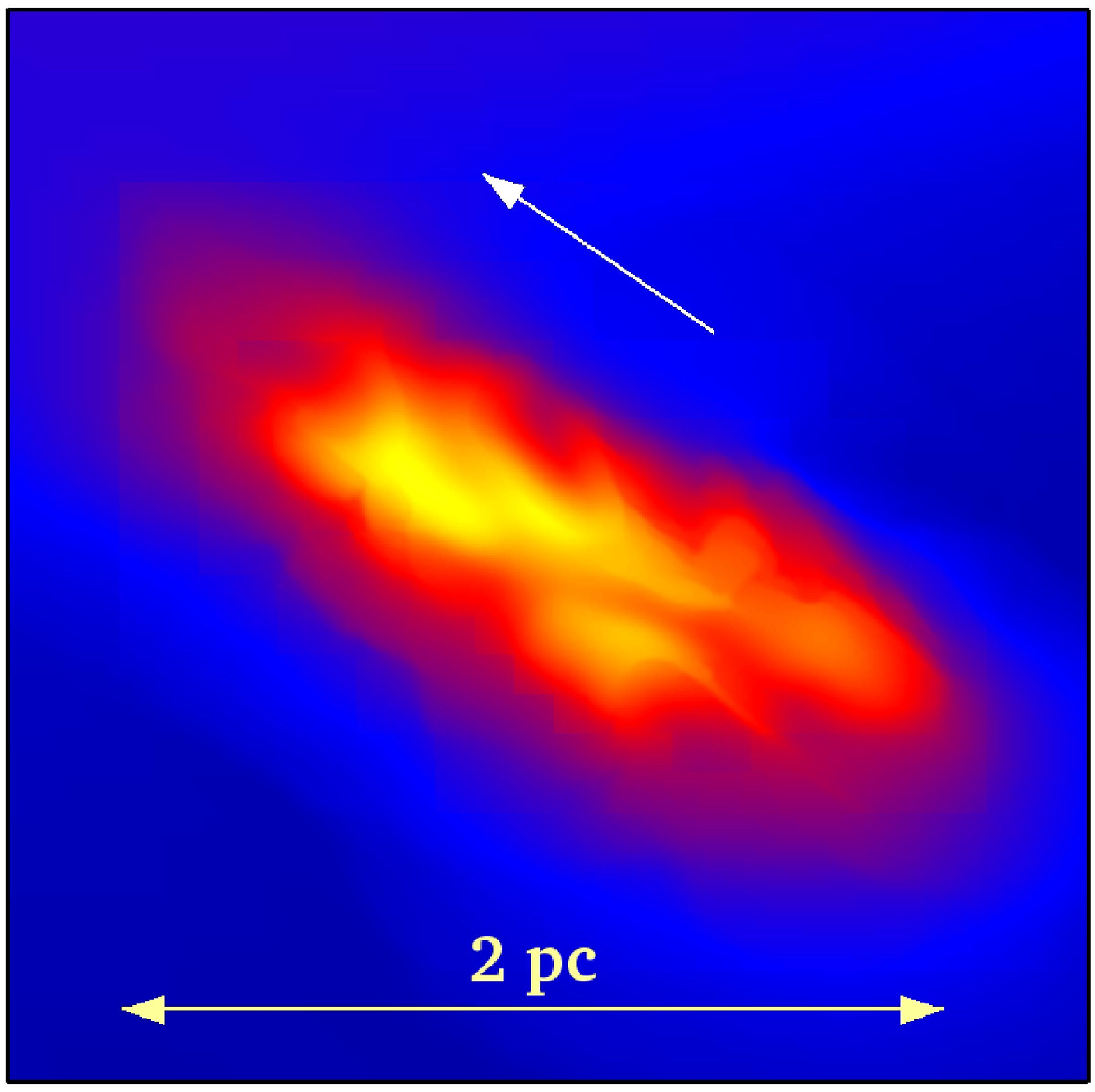}
\end{tabular}
\caption{Density slices through the center of the cloud at $\rm t=2/3t_{ff}$ for three different black hole mass models. The clouds are stretched in the direction of motion with increasing black hole mass, but compressed in the direction perpendicular to this. The black hole masses are: $\rm 10^{6} ~M_{\odot}$ for model M04 (left panel), $\rm 10^{7} ~M_{\odot}$ for model M01 (middle panel), and $\rm 10^{8} ~M_{\odot}$ for model M25 (right panel). The white arrow shows the direction of the orbital motion.}
\label{fig:morphology2}
\end{figure*}

From the phase diagrams we infer that the column density range is more extended for the lower shear runs. A salient feature is the dual phase structure that can be seen in the phase diagrams at low densities. This is because the cloud retains its size and shape for a longer period of time when there is less shear, developing a large column density range for the same densities. Moreover, we see from the density images that the cloud becomes more asymmetric with increasing $\rm M_{\rm bh}$. The cloud is stretched in the direction of the orbit but compressed perpendicular to it, see Fig. \ref{fig:morphology2}. Due to their morphological effect on the cloud, the gravitational stresses are the cause for the reduced column density range. This causes that in the presence of X-rays, a larger surface is directly irradiated by radiation with less attenuation throughout the cloud.

We find that the final star-formation efficiencies are reduced for higher black hole masses. Furthermore, we see that the star-formation rate has dropped as well, while the onset of star formation is delayed with increasing shear. However, when we look at the numbers of sink particles formed, we find that they are remarkably higher for large shear, that is, comparing $\rm M_{\rm bh}=10^{7} ~M_{\odot}$ against $\rm 10^{6} ~M_{\odot}$. The simulations with the highest shear, $\rm M_{\rm bh}=10^{8} ~M_{\odot}$, strongly inhibit the formation of all stars. See Table \ref{tab:table2} for details. This increase in the number of sink particles, while the SFE does not, shows that the formation of low-mass stars is favored for the case with $\rm M_{\rm bh}=10^{7} ~M_{\odot}$. When we look at the IMF plots, we do see that for the higher shear runs, the mass function generally comprises more low-mass stars and fewer high-mass stars. This result is in agreement with gravoturbulent fragmentation \citep{2004BaltA..13..365K}.

\subsection{Accretion onto sink particles}
Generally most of the mass growth of the sink particles comes from accretion. Proto-stars gain about 66\% of their mass in this way. This gives us an average accretion rate of $\sim$10$^{-5} \rm ~M_{\odot}/yr$. Due to the restricted time resolution of our checkpoint files, immediate merging of newly created sink particles within 1/30$^{\rm th}$ of a free-fall time ($\sim$3$\rm \times 10^{3} ~yr$), is considered as accretion. However, this should not occur very often, since only about 15\% of the particles merge during a run and this is spread throughout the whole simulation. The fraction of accreted mass at $\rm t=3t_{ff}$ is, on average, $\rm \bar{f}_{accr} \simeq 0.683$, when there is little black hole shear ($\rm M_{bh}=10^{6} ~M_{\odot}$). This reduces to $\rm \bar{f}_{accr} \simeq 0.642$ when we increase $\rm M_{\rm bh}$ to $\rm 10^{7} ~M_{\odot}$. Similarly, when we increase the X-ray flux from 0, 5.1, 28 to 160, the accreted mass fraction drops from 0.774, 0.746, 0.579 to 0.632 for $\rm M_{\rm bh}=10^{6} ~M_{\odot}$ and from 0.681, 0.652, 0.608 to 0.626 for $\rm M_{\rm bh}=10^{7} ~M_{\odot}$. See Table \ref{tab:table2} for a complete list of accreted mass fractions. Note that the highest X-ray flux has a bit higher mass accretion fraction than the second highest flux. This may be a consequence of the interplay between reduced accretion due to the increase in temperature, as sound speed scales with T$^{1/2}$, and the increase in accretion due to the larger gas reservoir available because of the lower star-formation efficiency. Apparently, the balance tips towards more accretion at the highest X-ray flux. The fractions for the runs with the highest shear are too volatile to consider, since very little star formation has occurred in these runs.

From Eq. \ref{eq:bondi-hoyle} we see that the accretion rate scales inversely with the sound speed and gas velocity to the third power. This is a strong scaling and is the dominant factor for the decrease in accretion as gas turbulence increases. The strong differential velocities due to shear from the black hole cascade down to smaller scales and yield turbulence. The turbulence still decays in time, but the decay is less rapid compared to when there is less shear. Thermal pressure as a consequence of the increased temperatures for the X-ray models also contributes to higher velocities and decreases the accretion rate. We do not see a clear correlation between accretion and cosmic rays, though. There can be two reasons for this. First of all, there is no gradient in pressure throughout the cloud due to the increase in cosmic ray ionization rate (they heat all gas uniformly). Second, the heating by cosmic rays is relatively modest, yielding much lower temperatures than that of UV or X-ray heated gas.

For the UV runs, we find notable differences after two free-fall times. Due to UV heating, accretion onto sink particles from the low density environment, $\rm n < 10^{4} ~cm^{-3}$, is drastically reduced. This is normally the main source for particles to gain mass at the later stages of the simulation. The more massive sink particles are also more strongly affected. Consequently, the models that produce a top-heavy IMF, these are the runs with increased X-ray flux, are restrained. A slightly more steep IMF slope is the result. The turn-over mass and the SFEs, on the other hand, are not significantly affected.

\section{Conclusions and discussion}
\label{sec:conclusions}
By means of numerical simulations, we have performed an extensive study on the effects of various kinds of irradiation (X-rays, cosmic rays, and UV) and shear on collapsing molecular clouds in active galactic environments. We have analyzed how the star-formation efficiencies and the initial mass functions are affected for 42 different cloud models. Our general result is that the IMF deviates from a Salpeter shape in extreme environments. Still, for a mixture of ambient conditions, their effects on the mass function somewhat balances out and a deviation from a Salpeter shape is less pronounced. We evaluate the differences and similarities in detail. All the quantifiable results are listed in Table \ref{tab:table2}.

The IMF nicely follows a lognormal distribution with a Salpeter slope for the isothermal runs. Our fiducial model M01 and model M04, which have similar conditions to those of the Milky Way, are in good agreement with a Chabrier IMF and match a Salpeter slope excellently, as can be seen in Fig. \ref{fig:imfs1}. Increasing the cosmic ray rate or the black hole shear does not change the slope of the mass function by much, except perhaps for the highest $\rm \zeta_{\rm CR}$ and the highest $\rm M_{\rm bh}$. Due to the insufficient number of sink particles, it is not possible to make a good fit for these cases.

The interplay between X-ray, shear and cosmic rays is as follows. The power-law slope of the IMF flattens and becomes non-Salpeter when we increase the X-ray flux from 0 to 160 $\rm erg ~s^{-1} ~cm^{-2}$. We find that the slope drops from around $\rm \Gamma_{\rm powfit} = -1.35$ ($\rm F_{X}=0$) to about $\rm \Gamma_{\rm powfit} = -1$ ($\rm F_{X}=160$) for the runs with $\rm M_{\rm bh}=10^{7} ~M_{\odot}$, and to about $\rm \Gamma_{\rm powfit} = -0.8$ ($\rm F_{X}=160$) when $\rm M_{\rm bh}=10^{6} ~M_{\odot}$. We see that the flattening is less prominent when the shear is stronger. We attribute this effect to the change in shape of the cloud, see Fig. \ref{fig:morphology2}. When gravitational shear is stronger, the irradiated face of the cloud is larger due to stretching of the cloud in the direction of the orbit. A bigger region is thus compressed by thermal pressure. We also see that the cloud obtains a rotational motion. Its rotational time is comparable to, though slightly faster than, the orbital time. This rotation is slow, however, it does cause the regions that were initially hot, to become shielded from X-rays. These regions, which had their densities raised by thermal compression now cool down quickly in the shade and form low-mass stars easily. This while the model clouds in the lowest $\rm M_{\rm bh}=10^{6} ~M_{\odot}$ runs face the radiation source on the same side for a much longer period, as their orbital time is about three times larger. This argument is supported by the behavior of the models with the highest shear. For these, the isothermal cosmic ray runs do not form sink particles at all, however, in the presence of X-rays, star formation does still occur thanks to compression. Cosmic rays, on the other hand, enhance the flattening of the power-law slope. At the highest $\rm \zeta_{CR}$, the IMF slope drops further from about $\rm \Gamma_{\rm powfit} =$-1 to -0.8 for the $\rm M_{bh}=10^{7} ~M_{\odot}$ runs, and from $\rm \Gamma_{\rm powfit} =$-0.8 to -0.6 for the $\rm M_{bh}=10^{6} ~M_{\odot}$ runs. We can understand this behavior since there are regions in the cloud that are shielded from X-rays and where the gas behaves in an isothermal manner. These places will form more massive stars because of the higher temperatures. Since we plot all the stars together in the IMF plot, a steep overall IMF slope is created.

We find that the turn-over mass shifts towards higher masses for increasing X-ray fluxes and cosmic ray rates. A simple Jeans mass argument is sufficient to explain this behavior. Since the Jeans mass strongly depends on the temperature (Eq. \ref{eq:jeans}), and this sets the minimum mass for the fragmentation scale, stars that form in massive cores tend to be more massive as well. The effect of cosmic rays is less prominent in the presence of X-rays due to the softer equation of state, as mentioned earlier. This, in the end, makes the turn-over mass less strongly dependent on the Jeans mass, and in part explains the constancy of the characteristic mass in many star-forming regions \citep{2008ApJ...681..365E, 2009MNRAS.392.1363B}.

X-rays reduce the final star-formation efficiency by up to 40\% in the presence of a strong X-ray flux, $\rm F_{X}=160 ~erg ~s^{-1} ~cm^{-2}$. This effect is very modest for the lowest X-ray flux, $\rm F_{X}=5.1 ~erg ~s^{-1} ~cm^{-2}$. In fact, there is a slight increase in the SFE for the lower cosmic ray rates. For X-rays to have any significant impact on to the efficiency of star formation, a flux of at least $\rm F_{X} \geq 5.1 ~erg ~s^{-1} ~cm^{-2}$ is needed. The main reason for the reduced efficiency is that the cloud evaporates from the irradiated side, thereby reducing the mass of the molecular cloud. This, however, also increases the densities at the same side due to an ionizing compression front such that the star-formation rate remains high.

\begin{figure*}[htb!]
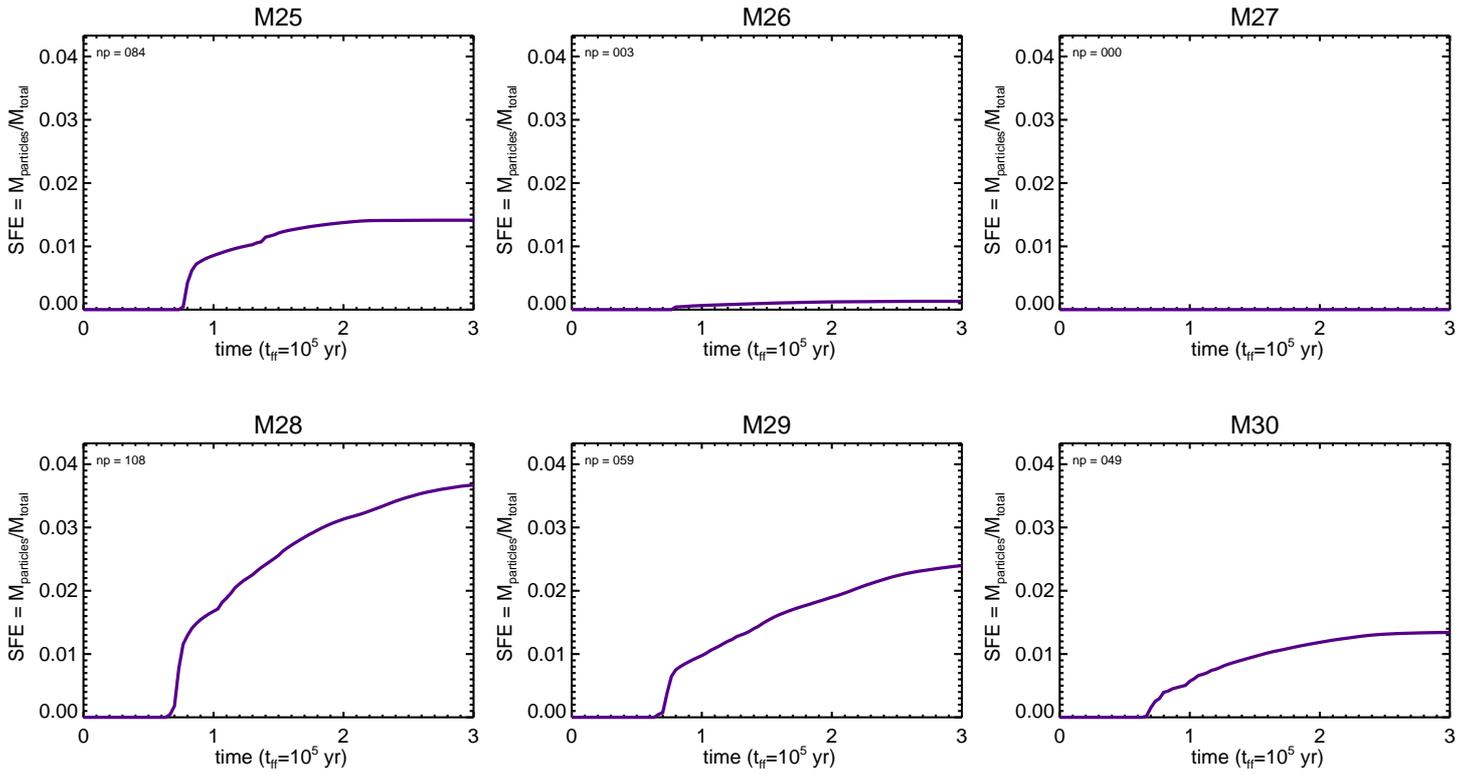

\hspace{-0.7cm}
\begin{tabular}{lll}

\vspace{0.7cm}

\begin{minipage}{6.0cm}
\includegraphics[scale=0.39]{sfe-m31-x0m8u0c1-lin.epsi}
\end{minipage} &
\begin{minipage}{6.0cm}
\includegraphics[scale=0.39]{sfe-m32-x0m8u0c100-lin.epsi}
\end{minipage} &
\begin{minipage}{6.0cm}
\includegraphics[scale=0.39]{sfe-m33-x0m8u0c3000-lin.epsi}
\end{minipage} \\

\begin{minipage}{6.0cm}
\includegraphics[scale=0.39]{sfe-m40-x160m8u0c1-lin.epsi}
\end{minipage} &
\begin{minipage}{6.0cm}
\includegraphics[scale=0.39]{sfe-m41-x160m8u0c100-lin.epsi}
\end{minipage} &
\begin{minipage}{6.0cm}
\includegraphics[scale=0.39]{sfe-m42-x160m8u0c3000-lin.epsi}
\end{minipage} \\

\end{tabular}
\caption{Star-formation efficiencies for the models with $\rm M_{\rm bh}=10^{8} ~M_{\odot}$ and UV=0. The ratio of the total sink particle mass over the total initial gas mass is plotted against time (in free-fall units). Note that the y-axis range in this figure differs from the other (SFE) figures. From left to right, the cosmic ray rate increases from 1, 100 to 3000 $\rm \times ~Galactic$. From top to bottom, the X-ray flux increases from 0 to 160 $\rm erg ~s^{-1} ~cm^{-2}$. The total number of sink particles formed during the run is given in the upper left corner of each panel.}
\label{fig:sfes3}
\end{figure*}

We also see a reduction in formation efficiency when we increase the black hole shear. We do this by increasing the black hole mass while keeping the distance of the cloud fixed at 10 pc, thereby changing the $\rm M_{\rm bh}/d_{\rm bh}^{3}$ ratio. For the runs with the highest shear, $\rm M_{bh}=10^{8} ~M_{\odot}$, star formation is almost completely quenched. In this case, the kinetic energy from the velocity divergence that the black hole injects is greater than the binding energy of the cloud. However, we find that stars can still form when thermal compression by X-rays enhances cloud collapse, albeit, with very low efficiency. The effect of shear on star formation is also more dramatic than that of the X-rays. Besides reducing the efficiency, the rate at which stars form is also affected. From the shape of the curves of the SFE plots in Figs. \ref{fig:sfes1} to \ref{fig:sfes-uv}, we see that the increase in SFE in time is less steep and somewhat more irregular for higher $\rm M_{\rm bh}$. The cause for this effect is the enhancement of the turbulence from the shearing motion cascading down to smaller scales. We see that the turbulence remains strong, FWHM = $\rm 4-5 ~km/s$ on the scale of the cloud at one free-fall time, for the $\rm M_{bh}=10^{7}$ and $\rm 10^{8} ~M_{\odot}$ runs, while the turbulence decays to about 1 km/s for $\rm M_{bh}=10^{6} ~M_{\odot}$. In addition to this, stars start to form later with increasing shear.

Cosmic rays also delay the onset of star formation. Interestingly, the star-formation efficiency and the star-formation rate are not affected. However, the number of sink particles is drastically reduced with increasing $\rm \zeta_{\rm CR}$. The average mass per sink particle, on the other hand, rises comparably, while the formation of low-mass stars is strongly inhibited. There is a significant side-effect from X-rays that counteracts this. When we look at the difference between the isothermal runs and the X-ray runs, for any $\rm \zeta_{\rm CR}$, we see that many more sink particles are formed when there is an X-ray source. Especially the number of low-mass stars is higher for the X-ray runs. This can be inferred form the shape of the mass function in Figs. \ref{fig:imfs1} to \ref{fig:imfs-uv}. One would normally expect that due to the increased gas temperatures from X-ray heating, for the same cosmic ray rate, star formation would be strongly suppressed. However, the equation of state plays a crucial role here. Despite the higher temperatures, the compressibility of the gas, $\rm \gamma$ being less than unity, causes the molecular cloud to fragment more easily. This while the cosmic ray runs enjoy an isothermal, $\rm \gamma=1$, equation of state. Therefore, low-mass stars are able to form with much less effort. The delay in star formation due to cosmic ray heating is also counterbalanced in the presence of X-rays. This effect is more prominent with increasing $\rm \zeta_{CR}$, see Fig. \ref{fig:multiplot1}. A time difference in the onset of star formation of more than one-third free-fall time ($\rm > 3.3\times10^{4} ~s$) can be seen for models with $\rm \zeta_{CR}=3000 \times Galactic$.

For our models with UV, we find that it has modest impact on the results. This was expected, since our model cloud has a column of $\rm N \simeq 10^{23} \rm ~cm^{-2}$ at the start of the simulation, and increases its (column) density with time. UV radiation is strongly attenuated above columns of $\rm N=10^{22} ~cm^{-2}$, for solar metallicity. We presented the models with UV for $\rm M_{bh}=10^{7} ~M_{\odot}$ and $\rm F_{UV} = 10^{2.5} ~G_{0}$ only, to asses whether increased external pressure and the higher cloud edge gas temperatures can play an important role in the evolution of the cloud. We find that by increasing the temperature for the lower densities, $\rm n<10^{4} ~cm^{-3}$, accretion onto particles is reduced at later times, i.e., $\rm t \gtrsim 2t_{ff}$. This impacts the growth of the massive particles later on. As such, the slope of the IMF, especially in the presence of X-rays, is somewhat more steep, in general. However, we find that the change in the SFEs are almost negligible when we have an isotropic UV field with a strength of $\rm 10^{2.5} ~G_{0}$, as compared to the non-UV runs. For cloud conditions such as in this work, we find that UV radiation can still be of importance, through secondary channels, to the formation of stars.

Accretion rates onto point particles in a turbulent medium may be different than the rates (Eq. \ref{eq:bondi-hoyle2}) used in this work as \cite{2006ApJ...638..369K} point out. Our medium is close to homogeneous on scales of a few parsec. On the other hand, outflows and jets from young stars could reduce the net accretion rates on smaller scales. Therefore, using a homogeneous flow approximation might overestimate the accretion rates and we consider them upper limits. In addition to this, \cite{2011ApJ...730...32S} show that episodic accretion for low-mass stars with bursts of radiative feedback can also affect the fragmentation properties. We have not taken such effects into consideration. We compare, however, our model results against each other. In this way, our findings are less sensitive to the above mentioned issues, and relative changes in the IMF are well defined.

The velocity shear through the cloud due to black hole gravity decreases as the cloud contracts. With time, the model cloud in our simulations shrinks in the direction perpendicular to the orbit due to self-gravity and turbulent motions. On average, the model cloud, shrinks by a factor of two, at most, in one free-fall time. This reduces the shear by the same amount as $\rm \triangle v$ scales with $\rm r_{cloud}$. Thus, the injected energy decreases with time, but not strongly.

In this study, the maximum resolution allowed was $4096^3$ cells. This in itself is quite a good resolution, and to follow it for long dynamical timescales, like we do, makes it difficult to go much higher. Since we use an adaptive grid, it is more important to properly resolve the grid dynamically. A proper means to do so is to resolve the Jeans length adequately everywhere in time and space. \cite{1997ApJ...489L.179T} say that the Jeans length should be resolved by at least 4 cells, but in the case of (M)HD simulations in magnetic fields, a minimum resolution of 30 cells is required according to \cite{2011ApJ...731...62F}. We resolved the Jeans length in our simulations by at least 10 cells. This actually means that the minimum resolution always lies somewhere between 10-20 cells. However, our resolution criterion is so strict that we resolve a whole block of 512 cells once even a single cell comes close to the minimum of 10 cells, while we only de-refine if all the cells within the block have stretched beyond 25 times the Jeans length. This in effect makes the general resolution in our runs much higher than the 10 cells that we mention, and approximately about 50-60 cells on average.

\begin{figure}[htb!]
\includegraphics[scale=0.54]{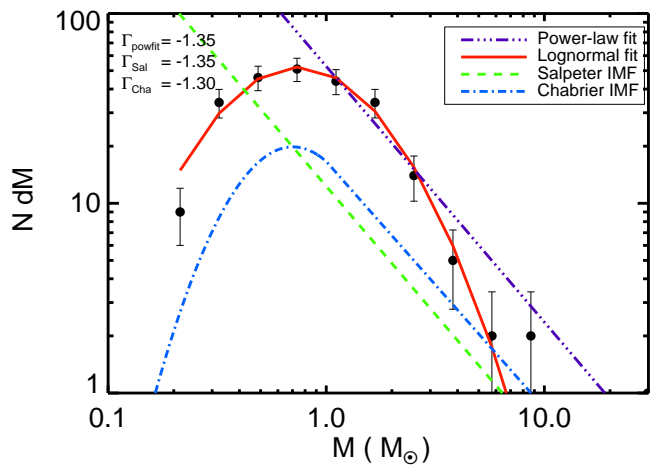}
\caption{The IMF of model M01 without sink particle merging. In this image, the Salpeter IMF (green dashed) and the Chabrier IMF (blue dot-dashed) are displayed as fitted to our fiducial model. A linear fit and a lognormal fit are shown as purple and red lines.}
\label{fig:m01nomerge}
\end{figure}

Stellar multiplicity is not covered in this work and binaries do not form naturally from the sink particle routine. In order to achieve that, cloud collapse should be followed to the highest grid resolutions in order to resolve the gas dynamics properly and allow proto-stars to form. This is very difficult to perform numerically. If we were to consider the sink particles in our work as binaries, then the found shapes of the IMFs would not be different, and would only be rescaled to lower masses. As a test, we have re-run 2 of our simulations without sink particle merging. We found that the merging routine does not influence the results much. Fig. \ref{fig:m01nomerge} shows a plot of model M01 without merging of sink particles.

In the future, we intend to perform studies for conditions much closer to the black hole, i.e., within 1 pc of galactic nuclei, to model massive stars as observed in Sgr A* and in M31. We also plan to incorporate our results into simulations of AGN evolution and intend to look into the stability of the entire disk.

\begin{acknowledgements}
We are grateful to the anonymous referee for an insightful and constructive report and the editor Malcolm Walmsley that helped to improve this work. We also would like to thank Dominik R.G. Schleicher for useful discussions on the behavior of the IMF in active galaxies. SH thanks M.A. Latif for insightful comments on the numerical part of this work and P. Papadopoulos for his eager input on cosmic rays. The FLASH code was in part developed by the DOE-supported Alliance Center for Astrophysical Thermonuclear Flashes (ACS) at the University of Chicago. The simulations have been run on the dedicated special purpose machines `Gemini' at the Kapteyn Astronomical Institute, University of Groningen and at the Donald Smits Center for Information Technology (CIT) using the Millipede Cluster. \\
\end{acknowledgements}

\bibliography{biblio.third.bib}

\clearpage

\clearpage

\begin{figure*}[htb!]
\hspace{-0.7cm}
\begin{tabular}{lll}

\vspace{0.7cm}

\begin{minipage}{6.0cm}
\includegraphics[scale=0.39]{sfe-m01-x0m7u0c1-lin.epsi}
\end{minipage} &
\begin{minipage}{6.0cm}
\includegraphics[scale=0.39]{sfe-m02-x0m7u0c100-lin.epsi}
\end{minipage} &
\begin{minipage}{6.0cm}
\includegraphics[scale=0.39]{sfe-m03-x0m7u0c3000-lin.epsi}
\end{minipage} \\

\vspace{0.7cm}

\begin{minipage}{6.0cm}
\includegraphics[scale=0.39]{sfe-m07-x5.1m7u0c1-lin.epsi}
\end{minipage} &
\begin{minipage}{6.0cm}
\includegraphics[scale=0.39]{sfe-m08-x5.1m7u0c100-lin.epsi}
\end{minipage} &
\begin{minipage}{6.0cm}
\includegraphics[scale=0.39]{sfe-m09-x5.1m7u0c3000-lin.epsi}
\end{minipage} \\

\vspace{0.7cm}

\begin{minipage}{6.0cm}
\includegraphics[scale=0.39]{sfe-m21-x28m7u0c1-lin.epsi}
\end{minipage} &
\begin{minipage}{6.0cm}
\includegraphics[scale=0.39]{sfe-m22-x28m7u0c100-lin.epsi}
\end{minipage} &
\begin{minipage}{6.0cm}
\includegraphics[scale=0.39]{sfe-m23-x28m7u0c3000-lin.epsi}
\end{minipage} \\

\begin{minipage}{6.0cm}
\includegraphics[scale=0.39]{sfe-m13-x160m7u0c1-lin.epsi}
\end{minipage} &
\begin{minipage}{6.0cm}
\includegraphics[scale=0.39]{sfe-m14-x160m7u0c100-lin.epsi}
\end{minipage} &
\begin{minipage}{6.0cm}
\includegraphics[scale=0.39]{sfe-m15-x160m7u0c3000-lin.epsi}
\end{minipage} \\

\end{tabular}
\caption{Star-formation efficiencies for the models with $\rm M_{\rm bh}=10^{7} ~M_{\odot}$ and UV=0. The ratio of the total sink particle mass over the total initial gas mass is plotted against time (in free-fall units). From left to right, the cosmic ray rate increases from 1, 100 to 3000 $\rm \times ~Galactic$. From top to bottom, the X-ray flux increases from 0, 5.1, 28 to 160 $\rm erg ~s^{-1} ~cm^{-2}$. The total number of sink particles formed during the run is given in the upper left corner of each panel.}
\label{fig:sfes1}
\end{figure*}

\begin{figure*}[htb!]
\hspace{-0.7cm}
\begin{tabular}{lll}

\vspace{0.7cm}

\begin{minipage}{6.0cm}
\includegraphics[scale=0.39]{sfe-m04-x0m6u0c1-lin.epsi}
\end{minipage} &
\begin{minipage}{6.0cm}
\includegraphics[scale=0.39]{sfe-m05-x0m6u0c100-lin.epsi}
\end{minipage} &
\begin{minipage}{6.0cm}
\includegraphics[scale=0.39]{sfe-m06-x0m6u0c3000-lin.epsi}
\end{minipage} \\

\vspace{0.7cm}

\begin{minipage}{6.0cm}
\includegraphics[scale=0.39]{sfe-m10-x5.1m6u0c1-lin.epsi}
\end{minipage} &
\begin{minipage}{6.0cm}
\includegraphics[scale=0.39]{sfe-m11-x5.1m6u0c100-lin.epsi}
\end{minipage} &
\begin{minipage}{6.0cm}
\includegraphics[scale=0.39]{sfe-m12-x5.1m6u0c3000-lin.epsi}
\end{minipage} \\

\vspace{0.7cm}

\begin{minipage}{6.0cm}
\includegraphics[scale=0.39]{sfe-m24-x28m6u0c1-lin.epsi}
\end{minipage} &
\begin{minipage}{6.0cm}
\includegraphics[scale=0.39]{sfe-m25-x28m6u0c100-lin.epsi}
\end{minipage} &
\begin{minipage}{6.0cm}
\includegraphics[scale=0.39]{sfe-m26-x28m6u0c3000-lin.epsi}
\end{minipage} \\

\begin{minipage}{6.0cm}
\includegraphics[scale=0.39]{sfe-m16-x160m6u0c1-lin.epsi}
\end{minipage} &
\begin{minipage}{6.0cm}
\includegraphics[scale=0.39]{sfe-m17-x160m6u0c100-lin.epsi}
\end{minipage} &
\begin{minipage}{6.0cm}
\includegraphics[scale=0.39]{sfe-m18-x160m6u0c3000-lin.epsi}
\end{minipage} \\

\end{tabular}
\caption{Star-formation efficiencies for the models with $\rm M_{\rm bh}=10^{6} ~M_{\odot}$ and UV=0. The ratio of the total sink particle mass over the total initial gas mass is plotted against time (in free-fall units). From left to right, the cosmic ray rate increases from 1, 100 to 3000 $\rm \times ~Galactic$. From top to bottom, the X-ray flux increases from 0, 5.1, 28 to 160 $\rm erg ~s^{-1} ~cm^{-2}$. The total number of sink particles formed during the run is given in the upper left corner of each panel.}
\label{fig:sfes2}
\end{figure*}

\begin{figure*}[htb!]
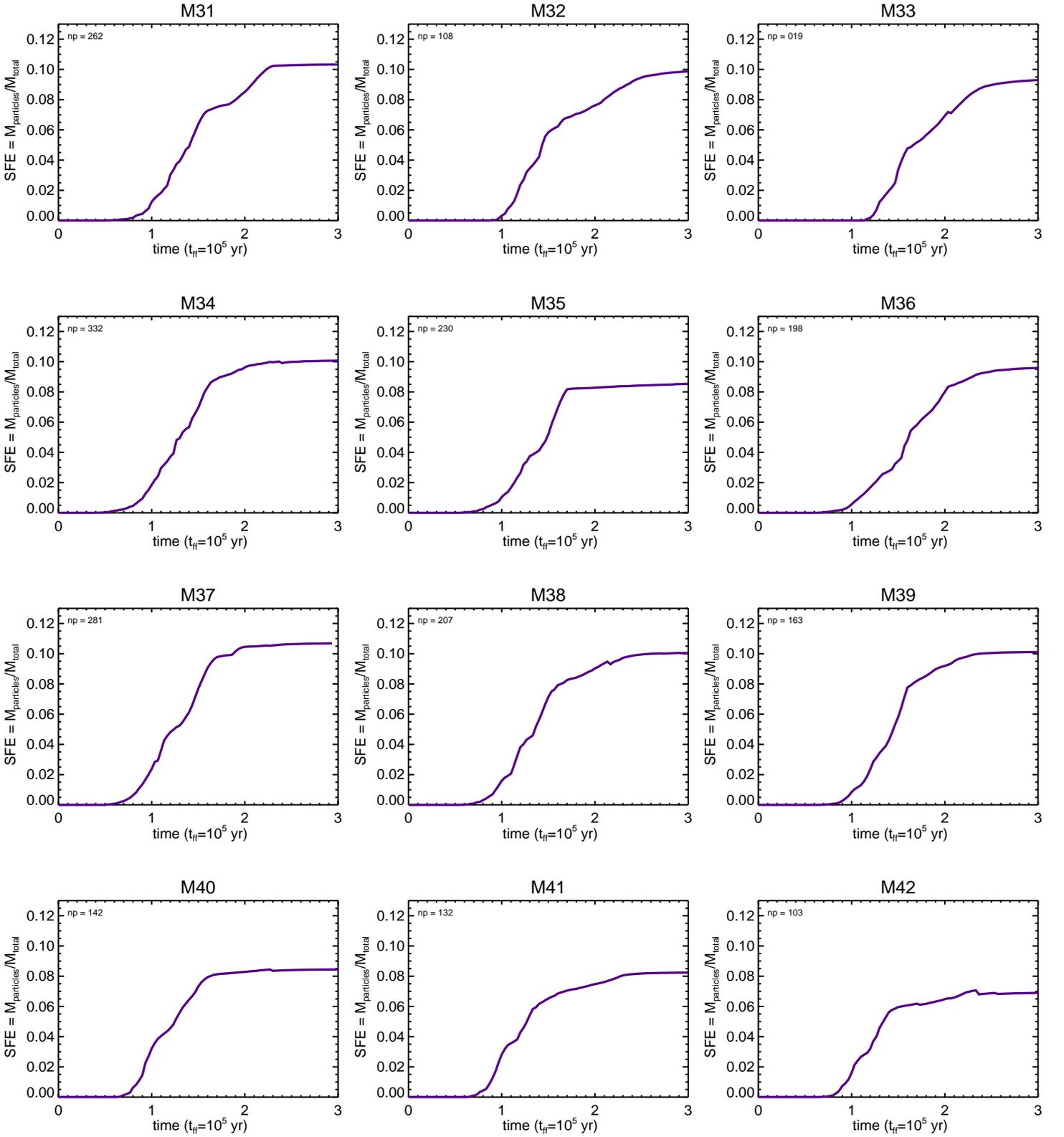

\hspace{-0.7cm}
\begin{tabular}{lll}

\vspace{0.7cm}

\begin{minipage}{6.0cm}
\includegraphics[scale=0.39]{sfe-m51-x0m7u2.5c1-lin.epsi}
\end{minipage} &
\begin{minipage}{6.0cm}
\includegraphics[scale=0.39]{sfe-m55-x0m7u2.5c100-lin.epsi}
\end{minipage} &
\begin{minipage}{6.0cm}
\includegraphics[scale=0.39]{sfe-m59-x0m7u2.5c3000-lin.epsi}
\end{minipage} \\

\vspace{0.7cm}

\begin{minipage}{6.0cm}
\includegraphics[scale=0.39]{sfe-m52-x5.1m7u2.5c1-lin.epsi}
\end{minipage} &
\begin{minipage}{6.0cm}
\includegraphics[scale=0.39]{sfe-m56-x5.1m7u2.5c100-lin.epsi}
\end{minipage} &
\begin{minipage}{6.0cm}
\includegraphics[scale=0.39]{sfe-m60-x5.1m7u2.5c3000-lin.epsi}
\end{minipage} \\

\vspace{0.7cm}

\begin{minipage}{6.0cm}
\includegraphics[scale=0.39]{sfe-m53-x28m7u2.5c1-lin.epsi}
\end{minipage} &
\begin{minipage}{6.0cm}
\includegraphics[scale=0.39]{sfe-m57-x28m7u2.5c100-lin.epsi}
\end{minipage} &
\begin{minipage}{6.0cm}
\includegraphics[scale=0.39]{sfe-m61-x28m7u2.5c3000-lin.epsi}
\end{minipage} \\

\begin{minipage}{6.0cm}
\includegraphics[scale=0.39]{sfe-m54-x160m7u2.5c1-lin.epsi}
\end{minipage} &
\begin{minipage}{6.0cm}
\includegraphics[scale=0.39]{sfe-m58-x160m7u2.5c100-lin.epsi}
\end{minipage} &
\begin{minipage}{6.0cm}
\includegraphics[scale=0.39]{sfe-m62-x160m7u2.5c3000-lin.epsi}
\end{minipage} \\

\end{tabular}
\caption{Star-formation efficiencies for the models with UV. The UV flux used in these models is $\rm 10^{2.5} ~G_{0}$. Similar to the figures before, Figs. \ref{fig:sfes1} to \ref{fig:sfes3}, the images show the ratio of the total sink particle mass over the total initial gas mass and is plotted against time (in free-fall units). From left to right, the cosmic ray rate increases from 1, 100 to 3000 $\rm \times ~Galactic$. From top to bottom, the X-ray flux increases from 0, 5.1, 28 to 160 $\rm erg ~s^{-1} ~cm^{-2}$. These UV models are simulated for a a black hole mass of $\rm M_{\rm bh}=10^{7} ~M_{\odot}$. The total number of sink particles formed during the run is given in the upper left corner of each panel.}
\label{fig:sfes-uv}
\end{figure*}

\clearpage

\begin{figure*}[htb!]
\hspace{-0.7cm}
\begin{tabular}{lll}

\vspace{0.7cm}

\begin{minipage}{6.0cm}
\includegraphics[scale=0.39]{m01-x0m7u0c1-90to30bins16.epsi}
\end{minipage} &
\begin{minipage}{6.0cm}
\includegraphics[scale=0.39]{m02-x0m7u0c100-90to30bins16.epsi}
\end{minipage} &
\begin{minipage}{6.0cm}
\includegraphics[scale=0.39]{m03-x0m7u0c3000-90to30bins16.epsi}
\end{minipage} \\

\vspace{0.7cm}

\begin{minipage}{6.0cm}
\includegraphics[scale=0.39]{m07-x5.1m7u0c1-90to30bins16.epsi}
\end{minipage} &
\begin{minipage}{6.0cm}
\includegraphics[scale=0.39]{m08-x5.1m7u0c100-90to30bins16.epsi}
\end{minipage} &
\begin{minipage}{6.0cm}
\includegraphics[scale=0.39]{m09-x5.1m7u0c3000-90to30bins16.epsi}
\end{minipage} \\

\vspace{0.7cm}

\begin{minipage}{6.0cm}
\includegraphics[scale=0.39]{m21-x28m7u0c1-90to30bins16.epsi}
\end{minipage} &
\begin{minipage}{6.0cm}
\includegraphics[scale=0.39]{m22-x28m7u0c100-90to30bins16.epsi}
\end{minipage} &
\begin{minipage}{6.0cm}
\includegraphics[scale=0.39]{m23-x28m7u0c3000-90to30bins16.epsi}
\end{minipage} \\

\begin{minipage}{6.0cm}
\includegraphics[scale=0.39]{m13-x160m7u0c1-90to30bins16.epsi}
\end{minipage} &
\begin{minipage}{6.0cm}
\includegraphics[scale=0.39]{m14-x160m7u0c100-90to30bins16.epsi}
\end{minipage} &
\begin{minipage}{6.0cm}
\includegraphics[scale=0.39]{m15-x160m7u0c3000-90to30bins16.epsi}
\end{minipage} \\

\end{tabular}
\caption{Initial mass functions for the models with $\rm M_{\rm bh}=10^{7} ~M_{\odot}$ and UV=0. The images display the time averaged IMFs between 1 and 3 free-fall times, where $\rm t_{ff}=10^{5} ~yr$. From left to right, the cosmic ray rate increases from 1, 100 to 3000 $\rm \times ~Galactic$. From top to bottom, the X-ray flux increases from 0, 5.1, 28 to 160 $\rm erg ~s^{-1} ~cm^{-2}$. In each image, for comparison purposes, the Salpeter IMF (green dashed) and the Chabrier IMF (blue dot-dashed) are displayed as fitted to our fiducial model. Two best fits are applied to the data, a linear fit and a lognormal fit, and are shown as purple and red lines. With the exception of the lognormal fit, the slopes above the turn-over mass are given in the upper left corner.}
\label{fig:imfs1}
\end{figure*}

\begin{figure*}[htb!]
\hspace{-0.7cm}
\begin{tabular}{lll}

\vspace{0.7cm}

\begin{minipage}{6.0cm}
\includegraphics[scale=0.39]{m04-x0m6u0c1-90to30bins16.epsi}
\end{minipage} &
\begin{minipage}{6.0cm}
\includegraphics[scale=0.39]{m05-x0m6u0c100-90to30bins16.epsi}
\end{minipage} &
\begin{minipage}{6.0cm}
\includegraphics[scale=0.39]{m06-x0m6u0c3000-90to30bins16.epsi}
\end{minipage} \\

\vspace{0.7cm}

\begin{minipage}{6.0cm}
\includegraphics[scale=0.39]{m10-x5.1m6u0c1-90to30bins16.epsi}
\end{minipage} &
\begin{minipage}{6.0cm}
\includegraphics[scale=0.39]{m11-x5.1m6u0c100-90to30bins16.epsi}
\end{minipage} &
\begin{minipage}{6.0cm}
\includegraphics[scale=0.39]{m12-x5.1m6u0c3000-90to30bins16.epsi}
\end{minipage} \\

\vspace{0.7cm}

\begin{minipage}{6.0cm}
\includegraphics[scale=0.39]{m24-x28m6u0c1-90to30bins16.epsi}
\end{minipage} &
\begin{minipage}{6.0cm}
\includegraphics[scale=0.39]{m25-x28m6u0c100-90to30bins16.epsi}
\end{minipage} &
\begin{minipage}{6.0cm}
\includegraphics[scale=0.39]{m26-x28m6u0c3000-90to30bins16.epsi}
\end{minipage} \\

\begin{minipage}{6.0cm}
\includegraphics[scale=0.39]{m16-x160m6u0c1-90to30bins16.epsi}
\end{minipage} &
\begin{minipage}{6.0cm}
\includegraphics[scale=0.39]{m17-x160m6u0c100-90to30bins16.epsi}
\end{minipage} &
\begin{minipage}{6.0cm}
\includegraphics[scale=0.39]{m18-x160m6u0c3000-90to30bins16.epsi}
\end{minipage} \\

\end{tabular}
\caption{Initial mass functions for the models with $\rm M_{\rm bh}=10^{6} ~M_{\odot}$ and UV=0. The images display the time averaged IMFs between 1 and 3 free-fall times, where $\rm t_{ff}=10^{5} ~yr$. From left to right, the cosmic ray rate increases from 1, 100 to 3000 $\rm \times ~Galactic$. From top to bottom, the X-ray flux increases from 0, 5.1, 28 to 160 $\rm erg ~s^{-1} ~cm^{-2}$. In each image, for comparison purposes, the Salpeter IMF (green dashed) and the Chabrier IMF (blue dot-dashed) are displayed as fitted to our fiducial model. Two best fits are applied to the data, a linear fit and a lognormal fit, and are shown as purple and red lines. With the exception of the lognormal fit, the slopes above the turn-over mass are given in the upper left corner.}
\label{fig:imfs2}
\end{figure*}

\begin{figure*}[htb!]
\hspace{-0.7cm}
\begin{tabular}{lll}

\vspace{0.7cm}

\begin{minipage}{6.0cm}
\includegraphics[scale=0.39]{m31-x0m8u0c1-90to30bins16.epsi}
\end{minipage} &
\begin{minipage}{6.0cm}
\includegraphics[scale=0.39]{m32-x0m8u0c100-90to30bins16.epsi}
\end{minipage} &
\begin{minipage}{6.0cm}
\includegraphics[scale=0.39]{m33-x0m8u0c3000-90to30bins16.epsi}
\end{minipage} \\

\begin{minipage}{6.0cm}
\includegraphics[scale=0.39]{m40-x160m8u0c1-90to30bins16.epsi}
\end{minipage} &
\begin{minipage}{6.0cm}
\includegraphics[scale=0.39]{m41-x160m8u0c100-90to30bins16.epsi}
\end{minipage} &
\begin{minipage}{6.0cm}
\includegraphics[scale=0.39]{m42-x160m8u0c3000-90to30bins16.epsi}
\end{minipage} \\

\end{tabular}
\caption{Initial mass functions for the models with $\rm M_{\rm bh}=10^{8} ~M_{\odot}$ and UV=0. The images display the time averaged IMFs between 1 and 3 free-fall times, where $\rm t_{ff}=10^{5} ~yr$. From left to right, the cosmic ray rate increases from 1, 100 to 3000 $\rm \times ~Galactic$. From top to bottom, the X-ray flux increases from 0 to 160 $\rm erg ~s^{-1} ~cm^{-2}$. In each image, for comparison purposes, the Salpeter IMF (green dashed) and the Chabrier IMF (blue dot-dashed) are displayed as fitted to our fiducial model. Two best fits are applied to the data, a linear fit and a lognormal fit, and are shown as purple and red lines. With the exception of the lognormal fit, the slopes above the turn-over mass are given in the upper left corner.}
\label{fig:imfs3}
\end{figure*}

\begin{figure*}[htb!]
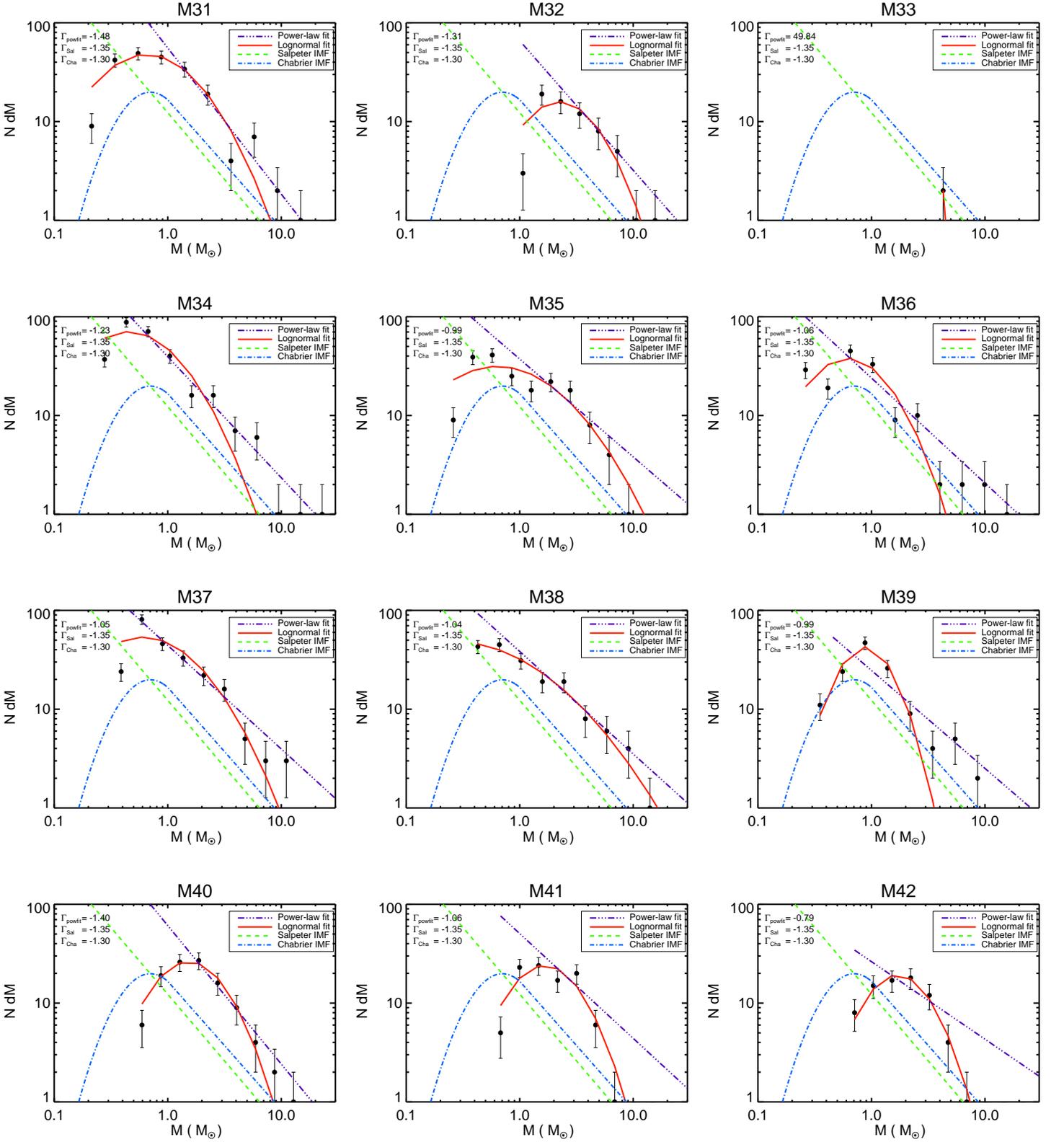

\hspace{-0.7cm}
\begin{tabular}{lll}

\vspace{0.7cm}

\begin{minipage}{6.0cm}
\includegraphics[scale=0.39]{m51-x0m7u2.5c1-90to30bins16.epsi}
\end{minipage} &
\begin{minipage}{6.0cm}
\includegraphics[scale=0.39]{m55-x0m7u2.5c100-90to30bins16.epsi}
\end{minipage} &
\begin{minipage}{6.0cm}
\includegraphics[scale=0.39]{m59-x0m7u2.5c3000-90to30bins16.epsi}
\end{minipage} \\

\vspace{0.7cm}

\begin{minipage}{6.0cm}
\includegraphics[scale=0.39]{m52-x5.1m7u2.5c1-90to30bins16.epsi}
\end{minipage} &
\begin{minipage}{6.0cm}
\includegraphics[scale=0.39]{m56-x5.1m7u2.5c100-90to30bins16.epsi}
\end{minipage} &
\begin{minipage}{6.0cm}
\includegraphics[scale=0.39]{m60-x5.1m7u2.5c3000-90to30bins16.epsi}
\end{minipage} \\

\vspace{0.7cm}

\begin{minipage}{6.0cm}
\includegraphics[scale=0.39]{m53-x28m7u2.5c1-90to30bins16.epsi}
\end{minipage} &
\begin{minipage}{6.0cm}
\includegraphics[scale=0.39]{m57-x28m7u2.5c100-90to30bins16.epsi}
\end{minipage} &
\begin{minipage}{6.0cm}
\includegraphics[scale=0.39]{m61-x28m7u2.5c3000-90to30bins16.epsi}
\end{minipage} \\

\begin{minipage}{6.0cm}
\includegraphics[scale=0.39]{m54-x160m7u2.5c1-90to30bins16.epsi}
\end{minipage} &
\begin{minipage}{6.0cm}
\includegraphics[scale=0.39]{m58-x160m7u2.5c100-90to30bins16.epsi}
\end{minipage} &
\begin{minipage}{6.0cm}
\includegraphics[scale=0.39]{m62-x160m7u2.5c3000-90to30bins16.epsi}
\end{minipage} \\

\end{tabular}
\caption{Initial mass functions for the models with UV. The UV flux used in these models is $\rm 10^{2.5} ~G_{0}$. Similar to the figures before, Figs. \ref{fig:imfs1} to \ref{fig:imfs3}, the images display the time averaged IMFs between 1 and 3 free-fall times, where $\rm t_{ff}=10^{5} ~yr$. From left to right, the cosmic ray rate increases from 1, 100 to 3000 $\rm \times ~Galactic$. From top to bottom, the X-ray flux increases from 0, 5.1, 28 to 160 $\rm erg ~s^{-1} ~cm^{-2}$. In each image, for comparison purposes, the Salpeter IMF (green dashed) and the Chabrier IMF (blue dot-dashed) are displayed as fitted to our fiducial model. Two best fits are applied to the data, a linear fit and a lognormal fit, and are shown as purple and red lines. With the exception of the lognormal fit, the slopes above the turn-over mass are given in the upper left corner. These UV models are simulated for a a black hole mass of $\rm M_{\rm bh}=10^{7} ~M_{\odot}$.}
\label{fig:imfs-uv}
\end{figure*}

\end{document}